\documentclass[aos]{imsart}

\RequirePackage[OT1]{fontenc}
\RequirePackage{amsthm,amsmath}
\RequirePackage[numbers]{natbib}
\RequirePackage[colorlinks,citecolor=blue,urlcolor=blue]{hyperref}

\startlocaldefs
\usepackage{pdfpages}

\usepackage{graphicx} 
\usepackage{bbm,amssymb}
\RequirePackage{multirow}

\usepackage{xr}
\usepackage{subfig}

\usepackage{color}
\definecolor{darkred}{RGB}{100,0,0}
\definecolor{darkgreen}{RGB}{0,100,0}
\definecolor{darkblue}{RGB}{0,0,150}

\usepackage{url}

\newtheorem{thm}{Theorem}[section]

\newtheorem{prp}[thm]{Proposition}
\newtheorem{lem}[thm]{Lemma}
\newtheorem{cor}[thm]{Corollary}

\numberwithin{equation}{section}
\theoremstyle{plain}

\def\beq{\begin{equation}}
\def\eeq{\end{equation}}
\def\beqn{\begin{eqnarray*}}
\def\eeqn{\end{eqnarray*}}
\def\bitem{\begin{itemize}}
\def\eitem{\end{itemize}}
\def\benum{\begin{enumerate}}
\def\eenum{\end{enumerate}}
\def\bmult{\begin{multline*}}
\def\emult{\end{multline*}}
\def\bcenter{\begin{center}}
\def\ecenter{\end{center}}

\DeclareMathOperator*{\argmax}{argmax}
\DeclareMathOperator*{\argmin}{argmin}

\DeclareMathOperator{\tr}{tr}

\def\bbK{\bbK}

\def\cP{\mathcal{P}}

\newcommand{\ac}[1]{\left\{#1\right\}}

\def\bA{\mathbf{A}}

\def\bE{\mathbf{E}}

\def\bX{\mathbf{X}}

\newcommand{\cro}[1]{\left[#1\right]}

\def\1{{\mathbf 1}}

\def\R{\mathbf{R}}

\def\bbK{\mathbb{K}}

\newcommand{\E}{\operatorname{\mathbb{E}}}
\renewcommand{\P}{\operatorname{\mathbb{P}}}

\newcommand{\Cov}{\operatorname{Cov}}

\newcommand{\MCOD}{\textsc{MCOD}}

\newcommand{\pa}[1]{\left(#1\right)}

\endlocaldefs

\begin{document}

\begin{frontmatter}

\title{ Model assisted variable clustering: minimax-optimal recovery and algorithms  }
\runtitle{ Variable clustering   } 

\begin{aug}
\author{\fnms{Florentina} \snm{Bunea}\ead[label=e2]{fb238@cornell.edu}}
\address{Department of Statistical Science\\
Cornell University\\
Ithaca, NY 14853-2601, USA\\
\printead{e2}}
\affiliation{Cornell University}

\author{\fnms{Christophe} \snm{Giraud}\ead[label=e1]{christophe.giraud@math.u-psud.fr}}
\address{Laboratoire de Math\'ematiques d'Orsay\\
Univ. Paris-Sud, CNRS, Universit\'e Paris-Saclay\\
F-91405 Orsay, France\\
\printead{e1}}
\affiliation{CNRS, Universit\'e Paris Sud, Universit\'e Paris-Saclay}

\author{\fnms{Xi} \snm{Luo}\ead[label=e6]{xi.rossi.luo@gmail.com}}
\address{Department of  Biostatistics\\
Brown University\\
Providence, RI 02903, USA\\
\printead{e6}}
\affiliation{Brown University}

\author{\fnms{Martin    } \snm{ Royer}\ead[label=e3]{martin.royer@math.u-psud.fr}}
\address{ Laboratoire de Math\'ematiques d'Orsay\\
Univ. Paris-Sud, CNRS, Universit\'e Paris-Saclay\\
F-91405 Orsay, France\\  
\printead{e3}}
\affiliation{CNRS, Universit\'e Paris Sud, Universit\'e Paris-Saclay  }

\and 
\author{\fnms{Nicolas   } \snm{Verzelen  }\ead[label=e4]{ nicolas.verzelen@inra.fr  }}
\address{  INRA, Montpellier SupAgro,  MISTEA\\  Univ. Montpellier\\ Montpellier, FRANCE
                            \printead{e4}}
\affiliation{INRA, Supagro, Univ. Montpellier }

\end{aug}




\begin{abstract}

The problem of variable clustering is  that of estimating groups of  similar components of a $p$-dimensional vector $X=(X_{1},\ldots,X_{p})$ from $n$ independent  copies of $X$.  
There exists a large number of algorithms that return data-dependent groups of variables, but their interpretation is limited to the algorithm that produced them. An alternative is model-based clustering, in which one begins by defining population level clusters  relative to a model that embeds notions of similarity. Algorithms tailored to such models 
yield estimated clusters with a clear statistical interpretation. We take this view here and introduce the class of $G$-block covariance models as a background model for variable clustering. In such models, two variables in a cluster are deemed similar if they have similar associations will all other variables. This can arise, for instance, when groups of variables 
are noise corrupted versions of the same latent factor. We quantify the difficulty of clustering data generated from a $G$-block covariance model in terms of cluster proximity, measured with respect to two related, but different,  cluster separation metrics.  We derive minimax cluster separation thresholds, which are  the metric values below which no algorithm can recover the model-defined clusters exactly, and show that they are  different for the two metrics.  We therefore develop two algorithms, COD and PECOK,  tailored  to $G$-block covariance models,  and study  their minimax-optimality with respect to each metric.  Of  independent interest is the fact that the analysis of the PECOK algorithm, which is based on a corrected convex relaxation of the popular $K$-means algorithm, provides the first statistical analysis of such  algorithms for variable clustering.  Additionally, we compare our methods with another popular clustering method, spectral clustering.  Extensive simulation studies, as well as our data analyses,  confirm the applicability of our approach.

\end{abstract}

\end{frontmatter} 


\section{Introduction}

The problem of variable clustering is that of grouping  similar components of a $p$-dimensional vector $X=(X_{1},\ldots,X_{p})$. These groups are referred to as clusters. In this work 
we investigate the problem of cluster recovery from a sample of $n$ independent copies of $X$. 
Variable clustering has had a long history in a variety of fields,  with important examples  stemming from  gene expression data \cite{GEM2Net, FreiditFrey2014, GeneExpressionCluster} or protein profile data  \cite{Bernardes2015}. The solutions to this problem are typically algorithmic and entirely data based. They include applications of $K$-means,  hierarchical clustering, spectral clustering, or versions of them.   The statistical 
properties of these procedures  have received a very limited amount of investigation.  It is not currently known what probabilistic cluster models on $X$ can be estimated by these popular techniques, or by their modifications. More generally, model-based variable clustering has received a limited amount of attention.  One net advantage of  model-based clustering is that population-level clusters are clearly defined, offering both interpretability of the clusters and a benchmark 
against which one can check the quality of a particular clustering algorithm.

In this work we propose the $G$-block covariance model as a flexible model for variable clustering and show that the clusters given by this model are uniquely defined. We then motivate and  develop two algorithms tailored to the model, COD and PECOK, and analyze their respective performance in terms of exact cluster recovery, for minimally separated clusters, under appropriately defined cluster separation metrics. 

\subsection{The $G$-block covariance model}

Our proposed  model  for variable clustering  subsumes that  the covariance matrix  $\Sigma$ of a centered random vector $X\in \R^p$  follows a block, or near-block,  decomposition, with blocks corresponding to a partition $G=\left\{G_{1},\ldots,G_{K}\right\}$ of $\{1,\ldots,p\}$. This structure of the covariance matrix has been observed to hold, empirically, in a number of very recent studies on the parcelation of the human brain, for instance  \cite{mert18, glasser16, craddock2012whole, yeo11}. We further support these findings in Section ~\ref{sec:data}, where we apply the clustering methods developed in this paper, tailored to $G$-block covariance models,  for the clustering of brain regions. \\

To describe our model,  we associate, to a partition $G$,  a membership matrix 
$A\in\R^{p\times K}$ defined by $A_{ak}=1$ if $a\in G_{k}$,  and $A_{ak}=0$ otherwise.

\medskip

\noindent  {\bf (A)  The exact $G$-block covariance model.} In view  of the above discussion, clustering the variables $(X_1,\ldots, X_p)$ amounts to find a minimal (i.e. coarsest partition) $G^*$, such that two variables belong to the same cluster  if they have the same covariance with all other variables. This implies that the covariance matrix $\Sigma$ of $X$ decomposes as 
\begin{equation}\label{modgen}
\Sigma = AC^* A^t + \Gamma, 
\end{equation}
where $A$ is relative to $G^*$, $C^*$ is a symmetric $K \times K$ matrix, and $\Gamma$ a diagonal matrix. When a such a decomposition exists with the partition $G^*$, we say that $X \in \R^p$ follows an (exact)  $G^*$-block covariance model.

{\bf  (i) $G$-Latent Model.} Such a  structure arises, for instance,  when components of $X$ that belong to the same group can be decomposed into the sum between a common latent variable and an uncorrelated random fluctuation.  Similarity within group is therefore given by  association with the same unobservable source. Specifically, the exact block-covariance model (\ref{modgen}) holds, with a diagonal matrix $\Gamma$,  when 
\begin{equation}\label{latent}  X_{a}=Z_{k(a)}+E_{a}, \end{equation} with $\Cov(Z_{k(a)},E_{a})=0$, $\Cov(Z)=C^*$, and the individual fluctuations $E_a$ are uncorrelated, and thus $E$ has diagonal covariance matrix $\Gamma$. 
The index assignment function  $k:\ac{1,\ldots,p}\to\ac{1,\ldots,K}$ is defined by $G_{k}=\ac{a:k(a)=k}$.  In practice, this model is used to justify the construction 
of a single variable that represents a cluster, the average of  $X_a$, $a \in G_k$, viewed as an observable proxy of $Z_{k(a)}$. 
For example,  a popular analysis approach for fMRI data,  called region-of-interest (ROI) analysis \cite{poldrack2007region},  requires averaging  the observations from multiple  voxels (a imaging unit for a small cubic volume of the brain)  within  each ROI (or cluster of voxels)  to produce new variables, each  representing a larger and interpretable brain area.  These new variables are then used for downstream analyses. From this perspective, model (\ref{latent}) can be used in practice, see, for example  \cite{bellec2006identification},  as a building block in a data analysis  based on cluster representatives,  which in turn  requires accurate cluster estimation. Indeed, data-driven methods for  clustering either voxels into regions or regions into functional systems, especially based on the  covariance matrix of $X$, is becoming increasingly important, see for example \cite{glasser16,yeo11,craddock2012whole,power2011functional}.  Accurate data-driven clustering methods also enable studying the cluster differences across subjects \cite{chong2017individual} or experimental conditions \cite{james2016human}.

\smallskip 

{\bf (ii) The Ising Block Model.} The Ising Block Model has been proposed in \cite{2016arXiv161203880B} for modelling 
social interactions, for instance political affinities.  Under this model, the joint distribution of  $X\in \{-1,1\}^p$, a $p$-dimensional vector with binary entries, is given by  
\beq\label{eq:density_ising}
 f (x) = \frac{1}{\kappa_{\alpha,\beta}}\exp\Big[\frac{\beta}{2p}\sum_{a\sim b}x_ax_b+ \frac{\alpha}{2p}\sum_{a\nsim b}x_ax_b\Big]\ ,  
\eeq
where the quantity $\kappa_{\alpha, \beta}$ is a normalizing constant, and the notation $a \sim b$ means that the elements are in the same group of the partition. The variables $X_a$  may for instance represent the votes of U.S. senators on a bill~\cite{banerjee2008model}. For parameters $\alpha>\beta$, the density~\eqref{eq:density_ising} models the fact that senators belonging to the same political group tend to share the same vote. By symmetry of the density $f$, the covariance matrix $\Sigma$ of $X$ decomposes as an exact block covariance model $\Sigma=A C^* A^t +\Gamma$ where $\Gamma$ is diagonal. When all groups $G^*_k$ have identical size, we have $C^*= (\omega_{in}-\omega_{out}) I_K+ \omega_{out}J$ and $\Gamma= (1-\omega_{in})I$, where  the $K \times K$ matrix $J$ has all entries equal to 1, and $I_K$ denotes the $K \times K$ identity matrix, and the quantities $\omega_{in},\omega_{out}$ depend on  $\alpha$,	  $\beta$, $p$.

\medskip

\noindent 
{\bf (B)  The approximate $G$-block model.} In many situations, it is more appealing to group variables that {\it nearly} share the same covariance with all the other variables.    In that situation, the covariance matrix $\Sigma$ would decompose as 
\begin{equation}\label{modgen_approx}
\Sigma = AC A^t + \Gamma, \text{ where }\Gamma\text{ has small  off-diagonal entries.}
\end{equation}
 Such a situation can arise, for instance when $X_{a}= (1 + \delta_a)Z_{k(a)}+ E_{a}$, with $\delta_a = o(1)$ and the individual fluctuations $E_{a}$ are uncorrelated, $ 1 \leq a \leq p$.

\subsection{Our contribution}
We assume that the data  consist in  i.i.d. observations $X^{(1)},\ldots,X^{(n)}$ of a random vector $X$ with mean 0 and covariance matrix $\Sigma$.
 This work  is 
devoted to the development of computationally feasible methods that yield estimates 
 $\widehat{G}$ of $G^*$, such that  $\widehat{G} = G^*$, with high probability, when the clusters are minimally separated, 
 and to characterize the minimal value of the cluster separation from a minimax perspective. 
 The separation between clusters  is a key element in quantifying the difficulty of a clustering task as,  intuitively,  well separated clusters should be easier to identify.   We consider two related, but different, separation metrics, that can be viewed as canonical whenever $\Sigma$ satisfies  (\ref{modgen_approx}).
 Although all our results allow, and are proved, for small departures from the diagonal structure of    $\Gamma$ in (\ref{modgen}),  our main contribution 
can be best seen when $\Gamma$ is a  diagonal matrix. We focus on this case below, for clarity of exposition. The case of $\Gamma$ being  a perturbation of a diagonal matrix is treated in Section \ref{sec:approx}.

When $\Gamma$ is diagonal, our target partition $G^*$ can be easily defined. It is the unique minimal (with respect to partition refinement) partition $G^*$
for which there is a decomposition $\Sigma=AC^*A^t+\Gamma$, with $A$ associated to $G^*$.
We refer to Section~\ref{sec:model} for details.  
We observe in particular, that $\max_{c\neq a,b}|\Sigma_{ac}-\Sigma_{bc}| > 0$ if and only if $X_a$ and $X_b$ belong to different clusters in $G^*$. 

This last remark motivates our first metric MCOD based on 
  the following  COvariance Difference (COD) measure  
 \begin{equation}\label{cod}\text{COD}(a, b)  :=  \max_{c\neq a,b}|\Sigma_{ac}-\Sigma_{bc}|\quad \textrm{for any}\ \ a,b=1,\ldots,p. \end{equation}
   We use the notation $a\stackrel{G^*}{\sim} b$ whenever $a$ and $b$ belong to the same group $G_k^*$, for some $k$, in the partition $G^*$, and similarly $a\stackrel{G^*}{\nsim} b$ means that there does not exist any group $G^*_k$ of the partition $G^*$ that contains both $a$ and $b$.
We   define  the  $\MCOD$ metric  as  \begin{equation}\label{mcod} \MCOD(\Sigma):=\min_{a\stackrel{G^*}{\nsim} b}\text{COD}(a,b).
 \end{equation} 
The measure $\text{COD}(a, b)$ quantifies the similarity of the  covariances  that $X_a$ and $X_b$ have, respectively, with all other variables.  From this perspective, the  size of  $\MCOD(\Sigma)$ is  a natural measure for the difficulty of clustering
 when analyzing clusters with components that are similar in this sense.  Moreover, note that this metric is well defined even if $C^*$ of   model (\ref{modgen}) is not semi-positive  definite.


 
 Another  cluster separation metric appears naturally when we view  model (\ref{modgen}) as arising via model (\ref{latent}), or via small deviations from it. Then, clusters  in  (\ref{modgen}) are driven by the latent factors, 
 and   intuitively they differ when the latent factors differ. Specifically, we define the "within-between group" covariance gap
\beq \label{definition_delta_distance}
\Delta(C^*):= \min_{j<k}\left(C^*_{kk}+C^*_{jj}-2C^*_{jk} \right)= \min_{j<k}\bE\left[(Z_{j}-Z_{k})^2\right], 
\eeq
where the second equality holds whenever (\ref{latent}) holds. In the latter  case, the matrix $C^*$, which is the covariance matrix of the latent factors, is necessarily semi-positive definite. Further, we observe that $\Delta(C^*) = 0$ implies $Z_j = Z_k$ a.s.  Conversely, we prove in Corollary \ref{du} of Section \ref{sec:model} that if the decomposition (\ref{modgen}) holds with $\Delta(C^*)>0$, then the partition related to $A$ is the partition  $G^*$ described above.
 An instance of $\Delta(C^*) > 0$  corresponds to having the within group covariances stronger than those between groups. This suggests the usage of this metric $\Delta(C^*)$ for cluster analysis  whenever, in addition to the general model formulation  (\ref{modgen}), we also expect clusters to have this property, which has been observed, empirically, to hold 
 in applications. For instance, it is implicit in the methods developed by  \cite{craddock2012whole}  for creating a human brain  atlas by partitioning appropriate covariance matrices.  
 We also  present a neuroscience-based data example in Section~\ref{sec:data}.   
 
 Formally, the two metrics are connected via the following chain of inequalities, proved in Lemma \ref{lem:equiv}  of Section \ref{sec:model} of the supplementary material \cite{SUPPL18}, and  valid as soon as the size of the smallest cluster is larger than one, $\Gamma$  and $C^*$ is semi-positive definite (for the last inequality)
 \begin{equation}\label{equiv} 2 \lambda_{K}(C^*) \leq   \Delta(C^*) \leq 2\MCOD(\Sigma) \leq 2 \sqrt{\Delta(C^*)}\ \max_{k=1,\ldots,K} \sqrt{C^*_{kk}}.\end{equation}
The first inequality shows that conditions on $\Delta(C^*)$ are weaker  than conditions on the minimal eigenvalue $\lambda_K(C^*)$ of $C^*$. In order to preserve the generality of our model, we do not necessarily assume that $ \lambda_{K}(C^*) > 0$, as we show that, for  model identifiability,  it is enough to have the weaker condition $\Delta(C^*)>0$, when the two quantities differ.

The second inequality in (\ref{equiv})  shows that $\Delta(C^*)$ and $\MCOD(\Sigma)$  can have the same order of magnitude, whereas the third inequality shows that they 
can also differ in order, and $\Delta(C^*)$ can be as small as $\MCOD^2(\Sigma)$, for small values of these metrics, which is our main focus. This suggests that different statistical assessments, and possibly different algorithms,  should be developed  for estimators  of clusters defined by (\ref{modgen}), depending on the  cluster separation metric.  To substantiate this intuition, we first derive, for each metric,  the rate below which no algorithm can recover exactly the clusters defined by (\ref{modgen}). We call this the minimax optimal threshold for cluster separation, and prove that it is different 
for the two metrics. We call an algorithm that can be proved to recover exactly  clusters with separation above the minimax threshold a minimax optimal algorithm.

 Theorem \ref{thm:tlbM} in Section \ref{SEC:MINIMAX}  shows  
that, for $K\geq 3$ and for some numerical constant $c>0$, no algorithm can estimate consistently clusters  defined by (\ref{modgen}) uniformly over covariance matrices fulfilling 
\begin{equation*}
\MCOD(\Sigma) \geq c  \sqrt{\log(p)\over n}. \end{equation*}
Theorem \ref{prp:minimax_lower_bound} in Section \ref{SEC:MINIMAX}  shows that optimal separation distances with respect to  the metric $\Delta(C^*)$ are sensitive to  the size of the smallest cluster,
\[ m^* = \min_{1 \leq k \leq K}|G^*_k|. \] Indeed, there exists a numerical constant $c>0$, such that  no algorithm can estimate consistently clusters defined by (\ref{modgen}) 
uniformly over covariance matrices fulfilling 
\begin{equation}\label{D} \Delta(C^*) \geq c   \left(\sqrt{\log(p)\over nm^*}\bigvee {\log(p)\over n}\right)\ .\end{equation}  
The first term will be dominant whenever the smallest cluster has size $m^* < n/\log (p)$, which will be the case in most situations.  The second  term in (\ref{D}) becomes dominant 
 whenever $m^* > n/\log (p)$, which can also happen when $p$ scales as $n$, and we have a few balanced clusters.

The PECOK algorithm is tailored to the $\Delta(C^*)$ metric, and is shown in Theorem \ref{THM:CONSISTENCY} to be near-minimax optimal. For instance, for balanced clusters, there exists a constant $c'$ such that exact recovery is guaranteed  when $\Delta(C^*) \geq c'\pa{ \sqrt{ \frac{K \vee  \log p} {m^*n}} + \frac{K \vee \log (p)}{n}}$.   This differs 
by factors in $K$ from the $\Delta(C^*)$-minimax threshold, for general $K$, whereas it is of optimal order when $K$ is a constant, or grows as slowly as $\log p$. 
A similar discrepancy  between minimax lower bounds and the performance of  polynomial-time estimators  has also been pinpointed in network clustering via the  stochastic block model~\cite{chen2014statistical} and in sparse PCA~\cite{pmlr-v30-Berthet13}.  It has been conjectured that, when  $K$ increases with $n$, there exists a gap between the statistical boundary, i.e. the minimal  cluster separation  for which a statistical method achieves perfect clustering with high probability, and the polynomial boundary, i.e. the minimal  cluster separation for which there exists a polynomial-time algorithm that  achieves  perfect clustering.  Further investigation of this computational trade-off is beyond  the scope of this paper and we refer to~\cite{chen2014statistical} and \cite{pmlr-v30-Berthet13} for more details. 

However, if we consider directly the  metric  $\MCOD({\Sigma})$,  and its corresponding, larger, minimax threshold, we derive the COD algorithm, which is  minimax optimal 
with respect to $\MCOD({\Sigma})$ when $K\geq 3$. In view of (\ref{equiv}), it is also minimax optimal with respect to $\Delta(C^*)$, whenever there exist small clusters, the size of which does not change with $n$. The  description of the two algorithms  and theoretical properties are given in Sections \ref{SEC:COD} and \ref{sec:pecok}, respectively, for exact block covariance models.  Companions of these results, regarding the performance of the algorithms for approximate block covariance models are given in Section \ref{sec:approx},  in Theorem \ref{prop:approxCOD} and Theorem \ref{prop:consistency-general-gamma-hat}, respectively.

Table \ref{table1} below gives a snap-shot of our results, which for ease of presentation,  correspond to the case of balanced clusters, with the same number of variables per cluster.  
We stress that neither our algorithms, nor our theory, is restricted to this case, but the exposition becomes more transparent in this situation. 

 \begin{table}[!h]
\centering
\begin{tabular}{|c||p{2cm}|p{3.2cm}|p{3.2cm}|}
\hline
  Metric & Minimax threshold & {\bf  PECOK}   &   {\bf   COD}   \\ \hline
 $d_1 = :\Delta(C^*)$ &$\sqrt{ \frac{  \log p} {mn}} + \frac{  \log p} {n}$ &  Minimax optimal w.r.t. $d_1$ when $K=O(\log(p))$.  & Minimax optimal w.r.t. $d_1$ when $m$ is constant. \\ \hline
 $d_2 =: \MCOD(\Sigma)$& $\sqrt{ \frac{  \log p} {n}}$ \ \ \ \ \ \ \ when $K\geq 3$& Minimax optimal w.r.t.  $d_2$ when  \ \ \ \ \ \  \ \ \ \  \ \ \ \ \ \ \  $m > n/\log(p)$ and $K = O(\log p)$.& Minimax optimal  w.r.t. $d_2$ when $K\geq 3$. \\ \hline
\end{tabular}
\caption{ Algorithm performance relative to  minimax thresholds of each metric\label{table1}} 
\end{table}

\noindent  In this table $m$ denotes the size of the smallest cluster in the partition. The performance of COD under $d_1$ follows from the second inequality in (\ref{equiv}), whereas the performance of PECOK under $d_2$ follows from the last inequality in (\ref{equiv}). The overall message transmitted by Table 1 and our  analysis is that, irrespective of the separation metric, the COD algorithm will be most powerful whenever we expect to have at least one, possibly more, small clusters, a situation that is typically not handled well in practice  by most of the popular clustering algorithms, see  \cite{bouveyron2014model} for an in-depth review.   The PECOK algorithm is expected to work best  for larger clusters, in particular when 
there are no clusters of size one.  We defer more comments on the relative numerical performance of the methods to the discussion Section \ref{practical}.

 We emphasize that both our algorithms are generally applicable, and our performance analysis is only  in terms of the most difficult scenarios, when two different clusters are almost indistinguishable and yet, as our results show, consistently estimable. Our extensive simulation results confirm these theoretical findings. \\

\noindent We summarize below our key  contributions. \\

\noindent  {\bf  (1) An identifiable model for variable clustering and metrics for cluster separation.}  
We advocate model-based variable clustering, as a way of proposing objectively defined  and interpretable clusters. 
We propose identifiable $G$-block covariance models for clustering, and prove cluster identifiability  in Proposition \ref{identif} of Section \ref{sec:model}.  

\noindent  {\bf  (2) Minimax lower bounds on cluster separation metrics for exact partition recovery.}
Two of our main results are Theorem \ref{prp:minimax_lower_bound} and Theorem \ref{thm:tlbM},  presented in Section~\ref{SEC:MINIMAX}, in which we  establish, respectively, minimax limits 
on the size of the $\Delta(C^*)$-cluster separation and   $\MCOD(\Sigma)$-cluster separation below which no algorithm can recover clusters defined by (\ref{modgen}) consistently, from a sample of size $n$ on $X$. To the best of our knowledge these are the first results of this type  in variable clustering. 

\noindent  {\bf  (3) Variable clustering procedures with guaranteed exact recovery  of minimally separated clusters.} 
The results of ({\bf 1}) and ({\bf 2}) provide a much needed  framework for motivating variable clustering  algorithm development 
and for  clustering algorithm assessments.  

In particular, they  motivate a correction of a convex relaxation of the $K$-means algorithm, leading to our proposed PECOK procedure, based on Semi-Definite Programing (SDP).  Theorem \ref{THM:CONSISTENCY}  shows it to be near-minimax optimal  with respect to the $\Delta(C^*)$ metric. The PECOK -  $\Delta(C^*)$ pairing is  natural, as $\Delta(C^*)$ measures the difference of the "within cluster" signal relative to  the "between clusters" signal, which is the idea that underlies $K$-means type procedures. To the best of our knowledge, this is the first work that  explicitly shows what model-based clusters of variables  can be estimated via $K$-means style methods, and assesses theoretically the quality of estimation.  Moreover, our work shows that the results obtained in ~\cite{2016arXiv161203880B}, for the block Ising model, can be generalized to arbitrary values of $K$ and unbalanced clusters. 

The COD procedure is a companion of PECOK for clusters given by model (\ref{modgen}), and is minimax optimal with respect to  the $\MCOD(\Sigma)$ cluster separation when $K\geq 3$, as established in Theorem \ref{thm:tlbM}.  Another advantage of COD is of computational nature, as SDP-based methods, although convex,  can be computationally involved.
 
  \noindent {\bf  (4)  Comparison with corrected spectral variable clustering methods.}  In Section \ref{SEC:SPECTRAL}, we  connect PECOK with another popular algorithm, spectral clustering. 
Spectral clustering is less computationally involved than PECOK, but the theoretical guaranties that we can offer for it are weaker.

\subsection{Organization of the paper} The rest of the  paper is organized as follows:  

Sections \ref{notation} and \ref{dis} contain the notation and distributional assumptions used throughout the paper. 

For clarity of exposition, Sections \ref{sec:model} - \ref{sec:pecok} contain results established for  model (\ref{modgen}), when  is $\Gamma$ a diagonal matrix. Extensions to the case when $\Gamma$ has small off-diagonal entries are presented in Section \ref{sec:approx}.

Section \ref{sec:model}  shows that we have a uniquely defined target of estimation, the partition $G^*$. 

Section \ref{SEC:MINIMAX}  derives the minimax thresholds on the separation metrics $\Delta(C^*)$ and  $MCOD(\Sigma)$, respectively, for estimating $G^*$ consistently. 

Section \ref{SEC:COD} is devoted to the COD algorithm, and its analysis. 

Section \ref{sec:pecok} is devoted to the PECOK algorithm and its analysis. 

Section \ref{SEC:SPECTRAL} analyses spectral clustering for variable clustering, and compares it with PECOK. 

Section \ref{sec:approx} contains extensions to approximate $G$-block covariance models. 

Section \ref{sec:data} presents their application to the clustering of putative brain areas  using a real fMRI data. 

Section \ref{sec:discussion}  contains a discussion of our results and overall  recommendations regarding the usage of our methods. 
Given the space constraints,  all proofs and simulation results are included  in the supplementary material. 

The implementation of PECOK can be found at \url{http://github.com/martinroyer/pecok/} 
and that of COD at \url{http://CRAN.R-project.org/package=cord}. 

\subsection{Notation}\label{notation}  
We denote by $\bX$ the $n\times p$   matrix  with rows corresponding to observations  $X^{(i)} \in \R^p$, for $i=1,\ldots,n$. The sample covariance matrix $\widehat{\Sigma}$ is defined by 
\[ \widehat{\Sigma} = \frac{1}{n} {\bf X}^{t}{\bf X}={1\over n}\sum_{i=1}^nX^{(i)}(X^{(i)})^t.  \]

Given a vector $v$ and $q\geq 1$, $|v|_{q}$ stands for the $\ell_q$ norm.
 For a generic matrix $M$:  $|M|_q$  denotes its  the entry-wise $\ell_q$ norm, $\|M\|_{op}$ denotes its operator norm, and $\|M \|_F$ refers to the Frobenius norm. We use $M_{:a} $, $M_{b:}$,  to denote the $a$-th column or, respectively,  $b$-th row of  a generic matrix $M$. 
The bracket $\langle.,.\rangle$ refers to the Frobenius scalar product. Given a matrix $M$, we denote $\mathrm{supp}(M)$ its support, that is the set of indices $(i,j)$ such that $M_{ij}\neq 0$.   $I$ denotes the identity matrix. We define the variation semi-norm of a diagonal matrix $D$ as $|D|_{V}:=\max_{a} D_{aa}-\min_{a} D_{aa}$.  We use $B \succcurlyeq 0$ to denote a symmetric and positive semidefinite matrix. \\

Throughout this paper will make use of the notation $c_1, c_2, \cdots $ to denote positive constants independent of $n, p, K, m$. The same letter, for instance $c_1$  may be used 
in different statements and may denote different constants, which are made clear within each statement, when there is no possibility for confusion. \\

We use $[p]$ to denote the set $\{1, \ldots, p\}$. We use the notation $a\stackrel{G}{\sim}b$ whenever $a, b \in G_{k}$, for the same $k$. Also,  $m=\min_k |G_k|$ stands for the size of the smallest group of the partition $G$.

The notation $\gtrsim$ and $\lesssim$ is used for whenever the inequalities hold up to multiplicative numerical constants.\subsection{Distributional assumptions} \label{dis}

For a $p$-dimensional random vector $Y$, its Orlicz norm is defined by $\|Y\|_{\psi_2}= \sup_{t\in \R^p:\  \|t\|_2=1}\inf\{s>0 :\ \mathbb{E}[e^{(Z^t t)/s^2}\leq 2] \}$. 
Throughout the paper we will assume that $X$ follows a sub-Gaussian distribution. Specifically, we use:

\indent {\bf Assumption 1}. (sub-Gaussian distributions) The exists $L>0$ such that  random vector $\Sigma^{-1/2}X$ satisfies  $\|\Sigma^{-1/2}X\|_{\psi_2}\leq L$, where 

Our class of distributions includes, in particular, that of bounded distributions, which may be of  independent interest, as   example (ii) illustrates.  We will therefore also specialize some of our results to this case, in which case we will use directly 

\indent {\bf Assumption 1-bis}. (Bounded distributions)  There exists $M>0$ such that  $\max_{i=1,\ldots,p} |X_i|\leq M$ almost surely.  

Gaussian distributions satisfy  Assumption 1  with $L=1$.  A bounded distribution is also sub-Gaussian, but the corresponding quantity $L$ can be much larger than $M$, and sharper results can be obtained if Assumption 1-bis holds. 

\section{Cluster identifiability in $G$-block  covariance models} \label{sec:model}

To keep the presentation focused, we consider in sections 2--5 the model (\ref{modgen}) with $\Gamma$   diagonal.  We treat the case corresponding to a diagonally dominant $\Gamma$ in Section \ref{sec:approx} below. In the sequel, it is assumed that $p>2$. 

We observe that  if the decomposition  (\ref{modgen}) holds  for a partition $G$, it also holds for any sub-partition of $G$. It is  natural therefore  to seek the smallest (coarsest) of such partitions, that is the partition with the least number of groups for which (\ref{modgen}) holds. Since the partition ordering is a partial order, the smallest partition is not necessarily unique.  However, the following lemma  shows that  uniqueness is guaranteed for our model class. 
\begin{lem}\label{lem:1_identif} Consider any  covariance matrix $\Sigma$. 
\begin{enumerate}
 \item[(a)] There exists a unique minimal partition $G^*$ such that $\Sigma= A C A^t + \Gamma$ for some diagonal matrix $\Gamma$, some membership matrix $A$ associated to $G^*$ and some matrix $C$. 
 \item[(b)]The partition  $G^*$ is given by the equivalence classes of the relation 
 \begin{equation}\label{def:COD}
 a\equiv b\ \ \textrm{if and only if}\ \ 
 COD(a,b):=\max_{c\neq a,b}|\Sigma_{ac}-\Sigma_{bc}|= 0.
 \end{equation}
\end{enumerate}
\end{lem}
\begin{proof}[Proof of Lemma \ref{lem:1_identif}]
If decomposition $\Sigma=ACA^t + \Gamma$ holds with $A$ related to a partition $G$, then we have $COD(a,b)=0$ for any $a,b$ belonging to the same group of $G$. Hence, each group $G_{k}$ of $G$ is included in one of the equivalence class of $\equiv$. As a consequence, $G$ is a finer partition than $G^*$ as defined in (b). 
Hence, $G^*$ is the (unique) minimal partition such that decomposition $\Sigma= ACA^t+\Gamma$ holds.
\end{proof}

As a consequence, the partition $G^*$ is well-defined and is identifiable. Next, we discuss the definitions of $\MCOD$ and $\Delta$ metrics. 
For any partition $G$, we let  $\MCOD(\Sigma, G):=\min_{a\stackrel{G}{\nsim} b}COD(a, b)$, where  we recall that the notation $a\stackrel{G}{\nsim} b$ means that $a$ and $b$ are not in a same group of  the partition $G$. By definition of $G^*$, we notice that $\MCOD(\Sigma,G^*)>0$ 
  and the  next proposition shows that $G^*$ is characterized by this property.
\begin{prp}\label{identif}
Let  $G$ be any partition such that $\MCOD(\Sigma,G) > 0$ and  the decomposition  $\Sigma=ACA^t+ \Gamma$ holds with $A$ associated to $G$.  Then $G=G^*$.
\end{prp}
The proofs of this proposition and the following corollary are given in Section \ref{app:identif} of the supplementary material \cite{SUPPL18}. In what follows, we use the notation 
 $\MCOD(\Sigma)$ for  $\MCOD(\Sigma, G^*)$. 
 
 In general, without further restrictions on the model parameters, the decomposition $\Sigma= AC A^t + \Gamma$ with $A$ relative to $G^*$ is not unique. If, for instance $\Sigma$ is the identity matrix $I$, then $G^*$ is the complete partition (with $p$ groups) and the decomposition \eqref{modgen} holds for any  $(C,\Gamma)=(\lambda I,(1-\lambda)I)$ with $\lambda\in \mathbf{R}$.  
 
 Recall that  $m^*:=\min |G^*_k|$ stands for the size of the smallest cluster. If we assume that  $m^*>1$ (no singleton), then $\Gamma$ is uniquely defined. Besides, the matrix  $C$ in \eqref{modgen} is only  defined up to a permutation of its rows and columns. In the sequel, we denote $C^*$ any of these matrices $C$.  When the partition contains  singletons ($m^*=1$), the matrix decomposition $\Sigma= AC A^t + \Gamma$ is made unique (up to a permutation of row and columns of $C$) by putting the additional constraint that the entries $\Gamma_{aa}$ corresponding to singletons are equal to 0. Since the definition of  $\Delta(C)$ is invariant with respect to permutation of rows and columns, this implies that $\Delta(C^*)$ is well-defined for any covariance matrix $\Sigma$.

For arbitrary $\Sigma$, $\Delta(C^*)$ is not necessarily positive. Nevertheless, if $\Delta(C^*)>0$, then $G^*$ is characterized by this property.
 
\begin{cor}\label{du}
Let $G$ be a partition such that $m=\min_{k}|G_{k}|\geq 2$, the decomposition  $\Sigma=ACA^t+ \Gamma$ holds with $A$ associated to $G$ and $\Delta(C)>0$. Then $G=G^*$. 
\end{cor}

As pointed in (\ref{definition_delta_distance}), in the latent model \eqref{latent}, $\Delta(C^*)$ is equal to the square of the minimal $L^2$-norm between two latent variables. So, in this case, the condition $\Delta(C^*)>0$ simply requires that all latent variables are distincts.

\section{Minimax thresholds on cluster separation for perfect recovery}\label{SEC:MINIMAX}

Before developing variable clustering procedures, we begin by assessing the limits of the size of each of the two cluster separation metrics  below which no algorithm can be expected to recover the clusters perfectly. 
We denote by $m^*=\min_{k}|G_{k}^*|$ the size of the smallest cluster of the target partition $G^*$ defined above. 
For $1\leq m\leq p/2$ and $\eta>0$, we define 
$ \mathcal{M}(m,\eta)$ as the set of covariance matrices $\Sigma$ fulfilling  $\text{MCOD}(\Sigma) > \eta|\Sigma|_{\infty}$ and whose associated partition $G^*$  has groups of equal size $m^*\geq m$.  Similarly, for $\tau>0$, we define $\mathcal{D}(m,\tau)$ as the set  of covariance matrices $\Sigma$ fulfilling $\Delta(C^*) > \tau |\Gamma|_{\infty}$  and whose associated partition $G^*$  has groups of equal size $m^*\geq m$. We use the notation $\P_{\Sigma}$  to refer to the normal distribution with covariance $\Sigma$.

\begin{thm}\label{thm:tlbM}
There exists a  positive constant $c_1$ such that, for any $1\leq m\leq p/3$ and any $\eta$ such that 
\begin{equation}\label{lbM}
0\leq \eta < \eta^{*}:= c_1\sqrt{\frac{ \log(p)}{n}} \,, 
\end{equation}
we have $\inf_{\widehat G}\sup_{\Sigma  \in  \mathcal{M}(m,\eta)} \mathbb{P}_{\Sigma}(\widehat G\neq G^{*})\geq 1/7$,
where the infimum is taken over all possible estimators. 
\smallskip

When $2\leq m=p/2$, the same result holds but with the Condition (\ref{lbM}) replaced by
\begin{equation}\label{lbM1}
 0 \leq \eta < \eta^*  := c_1 \left[\sqrt{\frac{\log(p)}{np}}\bigvee  \frac{\log(p)}{n} \right]. 
 \end{equation}
\end{thm} 

\noindent  We also have:

\begin{thm}\label{prp:minimax_lower_bound} 
There exist  positive constants $c_1$--$c_3$  such that the following holds for any $2\leq m\leq p/2$.  For any $\tau$ such that 
 \beq \label{eq:minimax_lower_bound}
 0 \leq \tau < \tau_*  := c_1 \left[\sqrt{\frac{\log(p)}{n(m-1)}}\bigvee  \frac{\log(p)}{n} \right],
\eeq
 then $  \inf_{\widehat G} \sup_{\Sigma \in \mathcal{D}(m,\tau)} \mathbb{P}_{\Sigma} \big[ \widehat{G}\neq G^*\big] \geq 1/7$,
where the infimum is taken over all estimators. 

Conversely, there exists a procedure $\widehat{G}$ satisfying $\sup_{\Sigma \in \mathcal{D}(m,\tau)} \mathbb{P}_{\Sigma} \big[ \widehat{G}\neq G^*\big] \leq c_3/p$ for any $\tau$ such that 
\[
 \tau > \tau^*  := c_2 \left[\sqrt{\frac{\log(p)}{n(m-1)}}\bigvee  \frac{\log(p)}{n}\right]
\]
\end{thm} 

Theorems  \ref{prp:minimax_lower_bound} and \ref{thm:tlbM} show that if either metric falls below the thresholds in (\ref{eq:minimax_lower_bound}) and (\ref{lbM}) or (\ref{lbM1}), respectively, the estimated partition $\widehat{G}$,  irrespective of the method of estimation, cannot achieve perfect recovery with high-probability uniformly over the set $ \mathcal{M}(m,\eta)$ or $\mathcal{D}(m,\tau)$. 
 The proofs are  given in Sections \ref{sec:proof:minimax}  of the supplement \cite{SUPPL18}. We note that the $\Delta(C^*)$ minimax threshold takes into account  the size $m^*$ of the smallest cluster, and therefore the required  cluster separation becomes smaller for large clusters. This is not the case for the second metric, as soon as there are at least 3 groups. The proof of  (\ref{lbM}) provides an example where we have $K = 3$ clusters,  that are very large, of size $m^* = p/3$ each, and where the $\MCOD(\Sigma)$ threshold  does not decrease with $m^*$.

 Theorem   \ref{prp:minimax_lower_bound} also provides a matching upper bound for the minimax threshold. Unfortunately, the procedure achieving this bound has an exponential computational complexity (see Section \ref{sec:proof:minimax3} in the supplementary material \cite{SUPPL18} for further details and Section \ref{sec:pecok} for a near-minimax optimal algorithm with polynomial computational complexity).

\section{COD for variable clustering}\label{SEC:COD}

\subsection{COD Procedure}
We begin with a procedure that can be viewed as natural  for model (\ref{modgen}).  It is based on the following intuition. 
Two indices  $a$ and $b$ belong to the same cluster of $G^*$, 
 if and only if  $\text{COD}(a, b) = 0$, with COD defined in (\ref{def:COD}). Equivalently, $a$ and $b$ belong to the same cluster when 
\[  s\text{COD}(a,b) =:  \max_{c \neq a, b}\frac{ |{\bf{cov}} (X_a - X_b, X_c)| } { \sqrt{{\bf var}(X_{b} -X_{a}){\bf var}(X_{c})} }= \max_{c \neq a, b} |{\bf cor}(X_{a}-X_{b},X_{c})| = 0, \] 
 where $s\text{COD}$ stands for {\it s}caled  COvariance Differences.  In the following we work with this quantity, as it is scale invariant. It is natural  to place $a$ and $b$ in the same cluster when the  estimator $\widehat{ s\text{COD}}(a,b)$  is below
 a certain threshold, where  
  \begin{equation}\label{cord}
 \widehat{s\rm COD}(a,b):= \max_{c\neq a,b} \left|\widehat{\bf cor}(X_{a}-X_{b},X_{c}) \right|=\max_{c\neq a,b} \left| {\widehat \Sigma_{ac}-\widehat \Sigma_{bc}\over \sqrt{\pa{\widehat \Sigma_{aa}+\widehat \Sigma_{bb}-2\widehat \Sigma_{ab}}{\widehat \Sigma_{cc}}}} \right|.
 \end{equation}

We estimate the partition  $\widehat G$ according to the simple  COD  algorithm explained  below.  The algorithm does not require as input the specification of the number $K$ of groups, which is automatically estimated by our procedure.    Step 3(c) of the algorithm is called the ``or" rule, and can be replaced with the ``and" rule below, without changing the theoretical properties of our algorithm, 
\[
{\widehat G_{l} = \ac{j \in S: \widehat{s\text{COD}}(a_{l},j)   \vee \widehat{s\text{COD}}(b_{l},j)\leq \alpha}}.
\] 
 The numerical performance of these two rules are also very close through simulation studies, same as we reported on a related COD procedure on correlations \cite{cord}. Due to these small differences, we will focus on the ``or" rule for the sake of space.
The algorithmic complexity for computing $\widehat{\Sigma}$  is $O(p^2n)$ and the complexity of COD  is $O(p^3)$, so the overall complexity of our estimation procedure is $O\big(p^2(p\vee n)\big)$. The procedure is also valid when $\Gamma$ has very small off-diagonal entries, and the results  are presented in Section \ref{sec:approx}.\\

\fbox{
\begin{minipage}
{0.96\textwidth}
{\bf The COD Algorithm} 
\begin{itemize}
\item Input:  $\widehat{\Sigma}$ and $\alpha>0$
\item Initialization: $S=\ac{1,\ldots,p}$ and $l=0$
\item Repeat: while $S\neq \emptyset$
\begin{enumerate}
\item $l \leftarrow l+1$
\item If $|S|=1$ Then $\widehat G_{l}=S$
\item If $|S|>1$ Then
\begin{enumerate}
\item $\displaystyle{(a_{l},b_{l})=\mathop{\mathrm{argmin}}_{a,b\in S,\ a\neq b} \widehat{s\text{COD}}(a,b)}$\smallskip
\item If $\displaystyle{ \widehat{s\text{COD}}(a_{l},b_{l})>\alpha}$ Then $\widehat G_{l} = \ac{a_{l}}$\\ \smallskip
Else $\displaystyle{\widehat G_{l} = \ac{j \in S: \widehat{s\text{COD}}(a_{l},j)\wedge \widehat{s\text{COD}}(b_{l},j)\leq \alpha}}$
\end{enumerate}
\item $S \leftarrow S\setminus \widehat G_{l}$
\end{enumerate}
\item Output: the partition $\widehat G=(\widehat G_{l})_{l=1,\ldots,k}$
\end{itemize}
\end{minipage}
}
\subsection{Perfect cluster recovery with COD for minimax optimal $\MCOD(\Sigma)$ cluster separation}

Theorem \ref{prop:COD} shows that the partition $\widehat{G}$ produced by the  COD algorithm has the property that  $\widehat G=G^*$, with high probability, 
as soon as  the separation $\MCOD(\Sigma)$ between clusters exceeds a constant times the threshold (\ref{lbM}) of Theorem \ref{thm:tlbM} of the previous section.

\begin{thm}\label{prop:COD}
Under the distributional Assumption 1, there exists numerical constants $c_{1},c_{2}>0$ such that,  
if  $$\alpha\geq c_{1} L^2 \sqrt{\log(p)\over n}$$ and
$MCOD (\Sigma) > 3\alpha |\Sigma|_{\infty}$, then we have exact cluster recovery with probability $1-c_{2}/p$.
\end{thm}
We recall that for Gaussian data, the constant $ L = 1$.  The proof is given in Section \ref{sec:proof:cod} of the supplement \cite{SUPPL18}. 

We observe that while the COD algorithm succeeds to recover $G^*$ at the minimax separation rate  (\ref{lbM}) when $K \geq 3$, it does not offer garanties at the minimax separation rate (\ref{lbM1}) when $K=2$. In this last case ($K=2$), we observe that 
$${1\over 2}\Delta(C^*) \leq \textrm{MCOD}(\Sigma) \leq \Delta(C^*),$$
so the metric MCOD$(\Sigma)$ is equivalent to $\Delta(C^*)$ and  we refer to the Section~\ref{sec:pecok} for an optimal algorithm.

\subsection{A Data-driven calibration procedure for COD}
\label{sec:cvcod}

The performance of the COD algorithm  depends on the value of the 
threshold parameter $\alpha$.  Whereas Theorem \ref{prop:COD} ensures that  a good  value for $\alpha$ is the order of  $\sqrt{\log p /n}$, its optimal value depends on the actual distribution (at least through the subGaussian norm) and is unknown to the statistician.  We propose below a new, fully data dependent,  criterion for selecting $\alpha$, and the corresponding partition $\widehat{G}$,  from a set of candidate partitions $\mathcal{G}$.  This criterion is based on data splitting. Let us consider two independent sub-samples of the original sample, $\mathcal{D}^{1}$ and $\mathcal{D}^{1}$,  each  of size $n/2$. 

We denote by $\widehat{\mathcal{G}}^{(1)}$ a collection of partitions computed from $\mathcal{D}^1$, for instance via  the COD algorithm with a varying threshold $\alpha$. 
For any $a<b$, we use $\mathcal{D}^{i}$, $ i = 1, 2$, to calculate, respectively
$$\widehat\Delta_{ab}^{(i)}=: \cro{\widehat{Cor}^{(i)}(X_{a}-X_{b},X_{c})}_{c\neq a,b} \ \  i = 1,2.$$

Since $\Delta_{ab}:=\cro{{Cor}(X_{a}-X_{b},X_{c})}_{c\neq a,b}$ equals zero if and only if $a \stackrel{G}{\sim}b$,
we want to select a partition $G$ 
such that $\widehat\Delta_{ab}^{(2)}{\bf 1}_{a \stackrel{G}{\nsim}b}$ is a good predictor of $\Delta_{ab}$. To implement this principle, it remains to evaluate $\Delta_{ab}$ independently of $\widehat\Delta_{ab}^{(2)}$. For this evaluation, we propose to re-use sample $\mathcal{D}^1$ which has already been used to build the family of partitions $\widehat{\mathcal{G}}^{(1)}$. More precisely, we select $\widehat G\in\widehat{\mathcal{G}}^{(1)}$  by minimizing the data-splitting criterion $\mathcal{H}$: 
$$\widehat G \in\mathop{\textrm{argmin}}_{G\in \widehat{\mathcal{G}}^{(1)}} \mathcal{H}(G)\quad {\rm with}\quad \mathcal{H}(G)=: \sum_{a<b}\cro{ | \widehat\Delta_{ab}^{(2)}{\bf 1}_{a \stackrel{G}{\nsim}b}-\widehat\Delta_{ab}^{(1)} |_{\infty}^2}.$$ 
\smallskip

The following proposition assesses the performance of $\widehat{G}$. We need the following additional assumption. 

${\bf (P1)}\quad$ If  $Cor(X_{a}-X_{b},X_{c})=0$ then $\E \widehat{Cor}(X_{a}-X_{b},X_{c})=0.$ \\
In general,  the sample correlation is not an unbiased estimator of  the population level correlation. Still, $({\bf P1})$ is satisfied when the data are normally distributed or in a latent model~\eqref{latent} when the noise  variables $E_{a}$ have a symmetric distribution. The next proposition provides guaranties for  criterion $\mathcal{H}$, averaged over $\mathcal{D}^2$, and denoted by  $\E^{(2)}[\mathcal{H}(G)]$.  The proof is given in Section \ref{sec:proof:COD2} of the supplement \cite{SUPPL18}. 
\begin{prp}\label{ECV}
Assume that the  distributional Assumption 1 and ({\bf P1}) hold. Then, there exists a constant $c_{1}>0$ such that, when
 $MCOD(\Sigma)>c_{1}|\Sigma |_{\infty}L^2\sqrt{\log(p)/n}$, we have 
\begin{equation}\label{eq:CVNEW}
\E^{(2)}[\mathcal{H} (G^{*})]\leq \min_{G\in\widehat{\mathcal{G}}^{(1)}}\E^{(2)}[\mathcal{H}(G)],
\end{equation}
both with probability larger than $1-4/p$  and in expectation with respect to $\P^{(1)}$.
\end{prp}
Under the condition $MCOD(\Sigma)>c_{1}|\Sigma |_{\infty}L^2\sqrt{\log(p)/n}$, Theorem  \ref{prop:COD} ensures that $G^*$ belongs to $\widehat{\mathcal{G}}^{(1)}$ with high probability, whereas \eqref{eq:CVNEW} suggests that the  criterion is minimized at $G^*$.

 If we  consider a data-splitting  algorithm based on  $\widehat{\text{COD}}(a, b)$ instead of $\widehat{\text{sCOD}}(a, b)$, then we can obtain a counterpart of Proposition \ref{ECV} without requiring the additional assumption  ${\bf (P1)}$. Still,  we favor the procedure based on  $\widehat{\text{sCOD}}(a, b)$ mainly for its scale-invariance property.

\section{ Penalized convex $K$-means: PECOK}\label{sec:pecok}

\subsection{PECOK Algorithm}
Motivated by the fact that  the COD algorithm is minimax optimal with respect to the $\MCOD(\Sigma)$ metric for $K\geq 3$, but not necessarily with respect to the $\Delta(C^*)$ metric (unless the size of the smallest cluster is constant), we propose below an alternative procedure, that adapts to this metric. Our second method is a natural extension of one of the most popular clustering strategies.  When we view the  $G$-block covariance model as arising via the latent factor representation in {\bf (i)}  in the Introduction, the canonical  clustering approach would be via the  $K$-means algorithm \cite{Lloyd}, which is NP-hard~\cite{awasthi2015hardness}.   Following  Peng and Wei \cite{PengWei07},  we consider a convex relaxation of it, which is computationally feasible in polynomial time. We argue below that, for  
estimating  clusters given by (\ref{modgen}),  one needs to further  tailor it to our model.  The statistical analysis of the modified procedure is the first to establish consistency of variable clustering via $K$-means type procedures, 
to the best of our knowledge.


The estimator offered  by the standard $K$-means algorithm, with the number $K$ of groups of $G^*$  known,  is 
\begin{equation}\label{def:critK}
 \widehat{G} \in\mathop{\textrm{argmin}}_{G}\text{crit}({\bf X},G)\quad \textrm{with}\quad \text{crit}({\bf X},G)=\sum_{a=1}^p\min_{k=1,\ldots,K}\|{\bf X}_{:a}-\bar {\bf X}_{G_{k}}\|^2,
 \end{equation}
and $\bar {\bf X}_{G_{k}}=|G_{k}|^{-1} \sum_{a\in G_{k}} {\bf X}_{:a}$.

For a partition $G$, let us introduce the corresponding  partnership matrix $B$ by 
\beq\label{B:matrix}
B_{ab}= \begin{cases} {1\over |G_{k}|} & \textrm{if $a$ and $b$ are in the same group $G_{k}$,}\\
0 & \textrm{if $a$ and $b$ are in a different groups}. 
\end{cases}
\eeq
we observe that $B_{ab}>0$ if and only if $a\stackrel{G}{\sim} b$.
In particular,  there is a one-to-one correspondence between partitions $G$ and their corresponding partnership matrices. It is shown in Peng and Wei~\cite{PengWei07} that the collection of such matrices $B$ is described by the collection $\mathcal{O}$ of orthogonal projectors fulfilling $\tr(B)=K$, $B1=1$ and $B_{ab}\geq 0$ for all $a,b$.

Theorem 2.2 in Peng and Wei \cite{PengWei07} shows that solving the $K$-means problem is equivalent to finding the global maximum 
\begin{equation}\label{KSDP1}
\bar B=\argmax_{B \in \mathcal{O}} \langle \widehat \Sigma,B\rangle\,, \end{equation}
and then recovering $\widehat{G}$ from $\bar{B}$.  

The set of orthogonal projectors is not convex, so, following Peng and Wei \cite{PengWei07}, we consider a convex relaxation $\mathcal{C}$ of $\mathcal{O}$ obtained by relaxing the condition "$B$ orthogonal projector", by "$B$ positive semi-definite", leading to
\beq\label{eq:domain}
\mathcal{C}:=\left\{ B \in \R^{p\times p}:
                \begin{array}{l}
                  \bullet\ B  \succcurlyeq 0 \  \ \text{(symmetric and positive semidefinite)} \\
                  \bullet\  \sum_a B_{ab} = 1,\ \forall b\\
		\bullet\ B_{ab}\geq 0,\ \forall a,b\\
		\bullet\ \tr(B) = K
                \end{array}
              \right\}.  
  \eeq
Thus, the (uncorrected) convex relaxation of $K$-means is equivalent with finding
\begin{equation}\label{km}
\widetilde B=\argmax_{B \in \mathcal{C}} \langle \widehat \Sigma,B\rangle .
\end{equation}

To assess the relevance of this estimator, we first study its behavior at the population level, when $\widehat{\Sigma}$ is replaced by $\Sigma$ in  (\ref{km}). Indeed, if the minimizer of our criterion  does not recover the true partition at the population level, we cannot expect it to be consistent, even in a large sample asymptotic context (fixed $p$, $n$ goes to infinity).  We recall that $|\Gamma|_V := \max_{a} \Gamma_{aa} - \min_{a} \Gamma_{aa}$. 

  \begin{prp}\label{prop:sdp-pop}
Assume that  $ \Delta(C^*)> {2|\Gamma|_{V}/ m^*}$. Then, $B^*=\argmax_{B\in\mathcal{O}} \langle \Sigma, B\rangle$. If $ \Delta(C^*)> {7|\Gamma|_{V}/ m^*}$, then  $B^*=\argmax_{B\in\mathcal{C}} \langle \Sigma, B\rangle$.
\end{prp}
For $\Delta(C^*)$  large enough, the population version of convexified $K$-means recovers $B^*$. The next proposition illustrates that the condition $ \Delta(C^*)> 2|\Gamma|_{V}/m^*$ for population $K$-means is in fact necessary.



 \begin{prp}\label{prp:counter_example} 
 Consider the model \eqref{modgen} with \[C^*=\left[{\scriptsize \begin{array}{ccc} 
    \alpha & 0 & 0\\
    0 & \beta & \beta-\tau\\
    0 & \beta-\tau & \beta
   \end{array}}\right]\ ,\quad  
\Gamma=\left[{\scriptsize\begin{array}{ccc}
\gamma_+&0&0\\ 0& \gamma_-&0 \\ 0& 0& \gamma_-
                 \end{array}}\right],\quad \text{ and \ }|G^*_1|=|G^*_2|=|G^*_3|=m^*\ .\]
The population maximizer  $B_{\Sigma}=\argmax_{B\in\mathcal{O}} \langle  \Sigma, B\rangle$ is not equal to $B^*$ as soon as 
$2\tau =\Delta(C^*)< \frac{2}{m^*}|\Gamma|_{V}$.
 \end{prp}
The two propositions above are proved in Section \ref{A1} in the supplementary material \cite{SUPPL18}. As a consequence, when  $\Gamma$ is not proportional to the identity matrix, the population minimizers  based on $K$-means and convexified $K$-means do not necessary recover the true partition even when the ''within-between group`` covariance gap is strictly positive. This undesirable behavior of $K$-means is not completely unexpected as $K$-means is a quantization algorithm which aims to find   clusters of similar width, instead of ''homogeneous'' clusters. Hence, we need to modify it for our purpose.

  This leads us to suggesting a  population level correction in Proposition \ref{prop:sdp-pop}.  Indeed, as a direct Corollary of Proposition \ref{prop:sdp-pop}, we have 
$$B^*=\argmin_{B\in\mathcal{C}} \langle \Sigma-\Gamma, B\rangle \ , $$
as long as $\Delta(C^*)> 0$.  This suggests the following {\bf Pe}nalized {\bf Co}nvex {\bf K}-means  (PECOK) algorithm,   in three steps.  The main step 2 produces an estimator $\widehat B$ of $B$ from which we derive the estimated partition $\widehat G$. We summarize this below.\\

\centerline{\fbox{  
\begin{minipage}{0.9\textwidth}
{\bf The PECOK algorithm}
\begin{align*} 
& \text{Step 1.  Estimate} \  \Gamma \ \text{by} \  \widehat{\Gamma}. \nonumber \\
& \text {Step 2.  Estimate} \ B^* \, \text{ by} \  
\widehat B=\argmax _{B \in \mathcal{C}} \left( \langle \widehat{\Sigma}, B\rangle  -    \langle \widehat{\Gamma}, B\rangle  \right). \nonumber \\ 
& \text{Step 3.  Estimate} \ G^* \, \text{ by  applying a clustering algorithm to the columns of }\,   \widehat{B}. \nonumber 
 \end{align*}
\end{minipage}
}}\medskip

The required inputs for Step 2 of our algorithm are:  
(i)  $\widehat{\Sigma}$,  
the sample covariance matrix; (ii) $\widehat{\Gamma}$, the estimator produced at Step 1; and (iii) $K$, the number of groups.   
Our only requirement  on the clustering algorithm applied in Step 3  is that it succeeds to recover the partition $G^*$ when applied to true partnership matrix $B^*$. The standard $K$-means algorithm \cite{Lloyd} seeded with $K$ distinct centroids, kmeans++ \cite{kmeans++}, or any  approximate $K$-means as defined in (\ref{eq:definition_eta_approximation}) in Section~\ref{SEC:SPECTRAL}, fulfill this property. 

 We view the term $   \langle \widehat{\Gamma}, B\rangle $ as a penalty term on $B$, with data dependent weights $\widehat \Gamma $. Therefore,  the construction of an  accurate estimator $\widehat \Gamma$ of $\Gamma$  is a  crucial step for guaranteeing the statistical optimality of  the PECOK estimator.

\subsection{Construction of \ $\widehat{\Gamma}$}

  Estimating $\Gamma$ before estimating the partition  itself is a non-trivial task, and needs to be done with care.  We explain our estimation below and analyze it in Proposition \ref{prp:control_u_gamma2} in Section~\ref{sec:proof_gamma_hat}. We show that this estimator of $\Gamma$ is appropriate whenever $\Gamma$ is a diagonal matrix (or diagonally dominant, with small off-diagonal entries). 
 For any $a,b\in [p]$, define 
\beq \label{eq:definition_V}
V(a,b):=   \max_{c,d \in [p]\setminus\{a,b\}} {\left| (\widehat \Sigma_{ac}-\widehat\Sigma_{ad})-(\widehat\Sigma_{bc}-\widehat\Sigma_{bd}) \right| \over \sqrt{\widehat \Sigma_{cc}+ \widehat \Sigma_{dd}-2 \widehat \Sigma_{cd}}}\ ,
\eeq
with the convention $0/0=0$.  Guided by the block structure of $\Sigma$, we define 
\[b_1(a):= \argmin_{b\in [p]\setminus\{a\}}V(a,b)\quad \text{ and }\quad b_2(a):= \argmin_{b\in [p]\setminus\{a,b_1(a)\}}V(a,b) ,\]
to be two  elements ''close'' to $a$, that is  two  indices  $b_1 = b_1(a)$ and $b_2 = b_2(a)$ 
such that the empirical covariance difference
$ \widehat \Sigma_{b_{i}c}- \widehat \Sigma_{b_{i}d}$, $i =1,2$,  is most similar to 
$ \widehat \Sigma_{ac}- \widehat \Sigma_{ad}$, for all variables $c$ and $d$ not equal to $a$ or $b_{i}$, $i = 1,2$.  It is expected that $b_1(a)$ and $b_2(a)$ either belong to the same group as $a$, or  belong to some ''close'' groups. 
Then, our estimator  $\widehat \Gamma$ is a diagonal matrix,  defined by 
\beq\label{eq:estim:gamma2}
\widehat \Gamma_{aa}=  \widehat \Sigma_{aa}+ \widehat \Sigma_{b_{1}(a)b_{2}(a)}- \widehat \Sigma_{ab_{1}(a)}- \widehat \Sigma_{ab_{2}(a)},
\quad \text{ for $a=1,\ldots, p$.}
\eeq 
Intuitively, $\widehat \Gamma_{aa}$ should be close to $\Sigma_{aa}+ \Sigma_{b_{1}(a)b_{2}(a)}- \Sigma_{ab_{1}(a)}-\Sigma_{ab_{2}(a)}$, which is equal to $\Gamma_{aa}$ in the favorable event where both $b_1(a)$ and $b_2(a)$ belong to the same group as $a$. 

In general, $b_1(a)$ and $b_2(a)$ cannot be guaranteed to belong to the same group as $a$. Nevertheless, these two surrogates $b_1(a)$ and $b_2(a)$  are close enough to $a$ so that  $|\widehat \Gamma_{aa}- \Gamma_{aa}|$  to be at most of the order  of $|\Gamma|_{\infty}\sqrt{\log(p)/n}$ in $\ell^\infty$-norm, as shown in Proposition \ref{prp:control_u_gamma2} in Section~\ref{sec:proof_gamma_hat} of the supplement material \cite{SUPPL18}. A slightly simpler estimator of $\Gamma$ was proposed in Appendix A of a previous version of this work ~\cite{pecock:old_arxiv}, but a bound on 
$|\widehat \Gamma_{aa}- \Gamma_{aa}|$ for that estimator contains a factor proportional to $|\Sigma|_{\infty}$, which is not desirable, and can be avoided by (\ref{eq:estim:gamma2}).
In the next subsection, we show that our proposed $\widehat{\Gamma}$ leads to perfect  recovery of  $G^*$, via PECOK,  under minimal separation conditions.

\smallskip
 
 Note that PECOK requires the knowledge of the true number $K$ of groups. 
 When the number $K$ of groups itself is unknown, we can modify the PECOK criterion by adding a penalty term as explained in a previous version of our work~\cite[Sec.~4]{pecock:old_arxiv}. Alternatively, we  propose  in Section \ref{sec:simu} of Supplement \cite{SUPPL18} selection via  a simple data-splitting procedure.


\subsection{Perfect cluster recovery with PECOK for near-minimax $\Delta$-cluster separation}\label{SEC:SUBSECT_PERFECT}

 
 We show in this section that the PECOK estimator  recovers the clusters exactly, with high probability, at a near-minimax 
 separation rate with respect to the $\Delta(C^*)$ metric.  
 
 \begin{thm}\label{THM:CONSISTENCY}
There exist $c_1,c_2, c_3$ three positive constants such that the following holds. 
Let $\widehat \Gamma $ be any estimator of $\Gamma$, such that 
$|\widehat \Gamma - \Gamma|_{V} \leq \delta_{n,p}$ with probability $1-c_{1}/p$. Then, under  Assumption 1, and when  
$L^4\log(p) \leq c_3 n$ and 
\begin{equation}\label{eq:condition_assumption2}
 \Delta(C^*) \geq c_L \left[\|\Gamma\|_{\infty}\left\{\sqrt{  \frac{\log p}{m^*n}   }+    \sqrt{\frac{p}{nm^{*2}}} + \frac{\log(p)}{n}+ \frac{p}{nm^*}\right\} +   \frac{\delta_{n,p}}{m^*} \right]\ ,
\end{equation}
then 
$\widehat{B} = B^*$ and $\widehat{G}=G^*$, with probability higher than $1 - c_1/p$. Here, $c_L$ is a positive constant that only depends on $L$ in Assumption 1. In particular, if $\widehat \Gamma$ is the estimator \eqref{eq:estim:gamma2},   the same conclusion holds with probability higher than $1-c_2/p$ when 
\begin{equation}\label{eq:gammaest}
 \Delta(C^*) \geq c_L \|\Gamma\|_{\infty}\left\{\sqrt{  \frac{\log p}{m^*n}   }+    \sqrt{\frac{p}{nm^{*2}}} + \frac{\log(p)}{n}+ \frac{p}{nm^*}\right\}. 
\end{equation}
\end{thm}

\noindent  The proof is given in Section \ref{T4.7} of the supplementary material \cite{SUPPL18}.

\noindent {\bf  Remark 1.} 
We left the term $\delta_{n,p}$ explicit in  (\ref{eq:condition_assumption2}) in order to make clear how 
the estimation of $\Gamma$ affects the cluster separation $\Delta(C^*)$ metric. Without a correction (i.e. taking $\widehat{\Gamma}=0$), the  term 
$\delta_{n, p}/m^*$ equals $|\Gamma|_{V}/m^*$ which is non zero (and does not decrease in a high-sample asymptotic) unless $\Gamma$ has equal diagonal entries. This phenomenon is consistent with the population analysis in the previous subsection.  Display (\ref{eq:gammaest}) shows that the separation condition can be much decreased with the correction.  In particular, for balanced clusters, that is when $m^* = \frac{p}{K}$,  exact recovery is guaranteed when 
 \beq\label{eq:condition_Delta_C}
 \Delta(C^*) \geq c_L\left[ \sqrt{ \frac{K \vee  \log p} {m^*n}} + \frac{K \vee  \log p} {n}\right]\ , 
 \eeq
for an appropriate constant $c_L > 0$. In view of Theorem \ref{prp:minimax_lower_bound}, when $m^*\geq c p/\log(p)$ the rate is minimax optimal, since in this case $K=p/m^*= O(\log(p))$. When $m^*=o(p/\log(p))$, the number $K$ of clusters  grows faster than $\log(p)$, and we possibly  lose a factor $K/\log(p)$ relative to the optimal rate.

As discussed in the introduction, this gap is possibly due to  a computational barrier and  we refer to~\cite{chen2014statistical} for a discussion in the related stochastic block model.\\

 Bounded variables $X$ also follow a sub-Gaussian distribution. Nevertheless, the corresponding sub-Gaussian norm $L$ may be large and Theorem \ref{THM:CONSISTENCY} can sometimes be improved, as in Theorem \ref{thm:consistencybounded} below, proved in Section \ref{T4.7}   of the supplementary material \cite{SUPPL18}.
  \begin{thm}\label{thm:consistencybounded}
There exist $c_1,c_2, c_3$ three positive constants such that the following holds. 
Let $\widehat \Gamma $ be any estimator of $\Gamma$, such that 
$|\widehat \Gamma - \Gamma|_{V} \leq \delta_{n,p}$ with probability $1-c_{1}/p$. Then, under  Assumption 1-bis, 
and 
\begin{eqnarray}\label{eq:condition_assumption1-bis}
 \Delta(C^*) &\geq& c_2 \Bigg[  M\|\Gamma\|_{\infty}^{1/2}\sqrt{\frac{p\log(p)}{nm^{*2}}}  +  M^2 \frac{p\log(p)}{nm^*}+ \frac{\delta_{n,p}}{m^*} \Bigg].
\end{eqnarray}
then 
$\widehat{B} = B^*$  and $\widehat{G}=G^*$, with probability higher than $1 - c_1/p$. 
\end{thm}
When we choose $\widehat{\Gamma}$ as in \eqref{eq:estim:gamma2}, the term $\delta_{n,p}/m^*$ can be simplified as under Assumption 1, see Proposition \ref{prp:control_u_gamma2} in Section \ref{sec:proof_gamma_hat} of the Supplement \cite{SUPPL18}. For balanced clusters, $m^* = \frac{p}{K}$, Condition \eqref{eq:condition_assumption1-bis} can be simplified in 
\[
 \Delta(C^*) \geq c_2 \Bigg[  M\|\Gamma\|_{\infty}^{1/2}\sqrt{\frac{K\log(p)}{nm^{*}}}  +  M^2 \frac{K\log(p)}{n}+ \frac{\delta_{n,p}}{m^*} \Bigg].
\]
In comparison to  \eqref{eq:condition_Delta_C}, the condition does no longer  depend on the sub-Gaussian norm $L$, but the term $K\vee \log(p)$ has been replaced by $K\log(p)$.

 \smallskip

\noindent {\bf Remark 2.}\label{ex:Ising}  For the Ising Block Model \eqref{eq:density_ising} with $K$ balanced groups,  we have $M = 1$ and $p=m^*K$, $C^*= (\omega_{in}-\omega_{out})I_K + \omega_{out}J$ and $\Gamma= (1-\omega_{in})I_K$. As a consequence,  no diagonal correction is needed, that is  we can take $\widehat{\Gamma}=0$,   and since $|\Gamma|_V = 0$, we have  $\delta_{n,p} = 0$. Then, for  $K$ balanced groups, condition (\ref{eq:condition_assumption1-bis})  simplifies to 
\[
 (\omega_{in}-\omega_{out}) \gtrsim   K\sqrt{\frac{\log(p)}{np}} + \frac{K\log(p)}{n}
\]
In the specific case $K=2$, we recover (up to numerical multiplicative constants) the optimal rate proved in~\cite{2016arXiv161203880B}. Our procedure and analysis provide a generalization of these results,  as they are  valid for general $K$ and Theorem \ref{thm:consistencybounded} also allows for unbalanced clusters.

\subsection{A comparison between PECOK and Spectral Clustering} \label{SEC:SPECTRAL}

In this section we discuss connections between the PECOK algorithm introduced above and  spectral clustering, a method 
that has  become  popular in  network clustering.  

First, we recall  the premise of spectral clustering, adapted to our context.  For $G^*$-block covariance models as \eqref{modgen},  we have $\Sigma - \Gamma = AC^*A^{t}$. Let $U$ be the $p \times K$ matrix collecting the  $K$ leading eigenvectors of $\Sigma - \Gamma$.  It has been shown, see e.g. Lemma 2.1 in Lei and Rinaldo \cite{LeiRinaldo},
that $a$ and $b$ belong to the same cluster if and  only if $U_{a:} = U_{b:}$ and  if and only if  $[UU^{t}]_{a:} =  [UU^{t}]_{b:}$.
When used for variable clustering, uncorrected spectral clustering  consists in  applying a clustering algorithm, such as $K$-means,  on the rows of the $p\times K$-matrix obtained by retaining  the $K$ leading eigenvectors of $\widehat \Sigma$.  
\medskip

\centerline{\fbox{
\begin{minipage}{0.9\textwidth}
{\bf SC algorithm}
\begin{enumerate}
\item Compute $\widehat{V}$, the matrix of the $K$ leading  eigenvectors of 
$ \widehat{\Sigma} $
\item Estimate $G^*$ by applying a (rotation invariant) clustering method to  the rows of $\widehat{V}$.  
\end{enumerate}
\end{minipage}
}}\medskip

 Arguing as in Peng and Wei~\cite{PengWei07}, we have the following. 
\begin{lem}\label{CSC}
SC algorithm 
 is equivalent to the following algorithm: 

\noindent Step 1.  Find  $
\overline{B}=\argmax\{ \langle  \widehat{\Sigma},B\rangle\ : \ tr(B)=K,\  I \succcurlyeq B \succcurlyeq 0\}$.

\noindent Step 2.  Estimate $G^*$  by applying a (rotation invariant) clustering method to  the rows of $\overline{B}$.
\end{lem} 

The connection between (unpenalized) PECOK and spectral clustering now becomes clear.
The (unpenalized) PECOK estimator $\widetilde B$ defined in \eqref{km} involves the calculation of 
\begin{equation}\label{SDP1bis}
\widetilde B=\argmax_{B}\{ \langle \widehat \Sigma,B\rangle \ : \ B1=1,\ B_{ab}\geq 0,\ tr(B)=K,\  B \succcurlyeq 0\}.
\end{equation}
Since the matrices $B$ involved in (\ref{SDP1bis}) are doubly stochastic, their eigenvalues are smaller than 1 and hence (\ref{SDP1bis}) is equivalent to 
$
\widetilde B=\argmax_{B}\{ \langle \widehat \Sigma,B\rangle\ : \ B1=1,\ B_{ab}\geq 0,\ tr(B)=K,\  I \succcurlyeq B \succcurlyeq 0\}$.
 Note then that 
$\overline{B}$ can be viewed as a less constrained version of $\widetilde{B}$, in which $\mathcal{C}$ is replaced by 
  $\overline{\mathcal{C}} = \{ B:  \ tr(B)=K,\  I \succcurlyeq B \succcurlyeq 0\}$,
 where we have dropped the  $p(p+1)/2$ constraints  given by  $B1=1$, and $B_{ab}\geq 0$. 
The proof of Lemma \ref{CSC} shows that $\overline{B}=\widehat{V}\widehat{V}^{t}$, so, contrary to $\widehat{B}$, the estimator $\overline{B}$ is (almost surely) never equal to $B^*$.
Below, we adapt the arguments of  \cite{LeiRinaldo}  in order to provide some guarantees for a corrected version of spectral clustering.
\medskip 

In view of this connection between Spectral clustering and unpenalized PECOK and of the fact that the population justification of spectral clustering deals with the spectral decomposition of $\Sigma -\Gamma$, we propose the following  corrected version of the algorithm, based on $\widetilde{\Sigma} := \widehat{\Sigma} - \widehat{\Gamma}$.

\medskip

\centerline{\fbox{
\begin{minipage}{0.9\textwidth}
{\bf CSC algorithm}
\begin{enumerate}
\item Compute $\widehat{U}$, the matrix of the $K$ leading  eigenvectors of 
$\widetilde{\Sigma} := \widehat{\Sigma} - \widehat{\Gamma}$
\item Estimate $G^*$ by clustering  the rows of $\widehat{U}$, via  an $\eta$-approximation of $K$-means (\ref{eq:definition_eta_approximation}). 
\end{enumerate}
\end{minipage}
}}\medskip

\noindent For $\eta>1$, an $\eta$-approximation of $K$-means  is a clustering algorithm producing a partition $\widehat G$ such that
\begin{equation}\label{eq:definition_eta_approximation}
\textrm{crit}\pa{\widehat U^t,\widehat G} \leq \eta\, \min_{G} \textrm{crit}\pa{\widehat U^t,G},
\end{equation}
with $\textrm{crit}(\cdot,\cdot)$ the $K$-means criterion (\ref{def:critK}).
Although solving $K$-means is NP-Hard~\cite{awasthi2015hardness}, there exist polynomial time approximate $K$-means algorithms, see  
Kumar {\it et al.}~\cite{ApproxKmeans}. As a consequence of the above discussion, the first step of CSC can be interpreted as a relaxation of the program associated to the PECOK estimator $\widehat{B}$.

\medskip

To simplify the presentation of the results for CSC procedure, we assume in the following that all the groups have the same size $|G^*_1|=\ldots=|G^*_K|=m^*=p/K$. We emphasize that this information is not required by either PECOK or CSC, or in the  proof of Proposition \ref{prp:spectral_clustering} below. We only 
use it here for simplicity.  We denote by $\mathcal{S}_{K}$  
the set of permutations on $\{1,\ldots,K\}$ and we denote by
$$\overline{L}(\widehat G,G^*)= \min_{\sigma\in \mathcal{S}_{K}}\sum_{k=1}^K{|G^*_{k}\setminus \widehat G_{\sigma(k)}|\over m^*}$$
 the sum of the ratios of miss-assigned  variables with indices in $G^*_k$.  In the previous sections, we studied perfect recovery of $G^*$, which would correspond to 
 $\overline{L}(\widehat G,G^*) = 0$. We give below conditions under which  $\overline{L}(\widehat G,G^*) \leq \rho$,  for an appropriate quantity  $\rho < 1$.  We begin with a general  theorem pertaining to partial partition recovery by CSC, under  a "signal-to-noise ratio" involving the  smallest eigenvalue $\lambda_{K}(C^*)$ of $C^*$.

\begin{prp}\label{prp:spectral_clustering}
Let  $Re(\Sigma)=tr(\Sigma)/\|\Sigma\|_{op}$ denote  the effective rank of $\Sigma$.
There exist  $c_{\eta,L}>0$ only depending  on $\eta$ and $L$ and a numerical constant $c_1$ such that the following holds under Assumption 1.
For any $0<\rho< 1$, if 
\begin{equation}\label{eq:spectral}
 \lambda_{K}(C^*)\geq {c_{\eta,L}\sqrt{K} \|\Sigma\|_{op}\over m^*\sqrt{\rho}}\sqrt{\frac{Re(\Sigma)\vee \log(p)}{n}},
\end{equation}
then $\overline{L}(\widehat G,G^*)\leq \rho$,  with probability larger than $1-c_1/p$.
\end{prp}

The proof extends  the arguments of \cite{LeiRinaldo},   initially developed for clustering procedures in stochastic block models, to our context. 
Specifically, we relate the error $\overline{L}(\widehat G,G^*)$ to the noise level, quantified in this problem by  $\|\widetilde{\Sigma}-AC^*A^t\|_{op}$.  We then employ the results of 
 \cite{koltchinskii2014concentration} to show that this 
operator norm can be controlled, with high probability, which leads to the conclusion of the theorem.  \smallskip

As $n$ goes to infinity, the right hand side of Condition \eqref{eq:spectral} goes to zero, and CSC is therefore consistent in a large sample asymptotic. In contrast, we emphasize that   (uncorrected) SC algorithm is  not consistent as can be shown by a population analysis similar to that of Proposition \ref{prp:counter_example}. \smallskip

We observe that $\Delta(C^*)\geq 2 \lambda_{K}(C^*)$, so we can compare the lower bound  (\ref{eq:spectral}) on $ \lambda_{K}(C^*)$ to the lower-bound (\ref{eq:condition_Delta_C}) on $\Delta(C^*)$. 
To further facilitate the comparison between CSC  and PECOK,  we discuss both the conditions and the conclusion of 
this theorem  in the simple setting where $C^*=\tau I_K$ and $\Gamma=I_p$. Then, the cluster separation measures coincide up to a factor 2,  $\Delta (C^*) = 2\lambda_K(C^*) = 2\tau$.

\begin{cor}[Illustrative example: $C^*=\tau I_K$ and $\Gamma=I_p$]\label{cor1_spectral} There exist three positive numerical constants $c_{\eta,L}$, $c'_{\eta,L}$ and $c_3$ such that the following holds under Assumption 1.
 For any $0<\rho<1$, if 
\beq\label{eq:condition_consistency_low_rank}
 \rho\geq c_{\eta,L}\Big[ \frac{K^2}{n}+ \frac{K\log(p)}{n}\Big]\quad \quad \text{ and }\quad \quad \tau \geq c'_{\eta,L}\Big[ {K^2\over \rho n} \vee \frac{K}{\sqrt{\rho nm^*}}\Big]\ ,
 \eeq
then $\overline{L}(\widehat G,G^*)\leq \rho$, with probability larger than $1-c_3/p$.~\\
\end{cor}

Recall that, Theorem  \ref{THM:CONSISTENCY} above states that, when $\widehat{G}$ is obtained via the PECOK algorithm, and  if 
$ \tau \gtrsim \sqrt{  \frac{K\vee \log p }{m^*n}   }+     \frac{\log(p)\vee K }{n}\ $, then  $\overline{L}(\widehat G,G^*) = 0$, or equivalently, $\widehat{G} = G^*$, with high probability. We can  therefore provide the following comparison (we refer to Section \ref{sec:simu} of the supplementary material \cite{SUPPL18} for a numerical comparison).
\begin{itemize}
\item when $ \tau \gtrsim \sqrt{  \frac{K\vee \log p }{m^*n}   }+     \frac{\log(p)\vee K }{n}\ $, and under the additional condition that $n \gtrsim (K\vee \log(p))^2/K$, CSC algorithm satisfies  $\overline{L}(\widehat G,G^*)\leq K^2/(K\vee \log(p))$. So, for $K=o({\log(p)})$ and for a large enough sample size $n \gtrsim (\log(p))^2/K$, the fraction of misclassified variables by CSC is vanishing as $O(K/\log(p))$ for $ \tau \gtrsim \sqrt{  \frac{\log p }{m^*n}   }+     \frac{\log(p) }{n}$.  This guaranty is slightly weaker than for PECOK which ensures exact recovery in this setting. This discrepancy may be an artifact of the proof technique. Very recent works \cite{abbe2017entrywise,lu2016statistical} (released during the reviewing process of this paper) present reconstruction error bounds tighter than those of \cite{LeiRinaldo},  for (variants of) spectral clustering, when applied  to  two parameter SBMs, for  network data, not the type of data analyzed in this work. 
\item When we move away from  the case $C^*=\tau I_K$, the guaranties for CSC can degenerate.  For instance, when $\Gamma=I$ and   $C^*=\tau I_K+\alpha J$, with $J$ being the matrix with all entries equal to one, as in the Ising Block model discussed page \pageref{ex:Ising}. Notice  that in this case we continue to have $\Delta(C^*)= 2\lambda_{K}(C^*)= 2\tau$. Then, for a given, fixed, value of $\rho$ 
 and $K$ fixed, condition \eqref{eq:spectral} requires a cluster separation at least
\[\tau \gtrsim \frac{\alpha\sqrt{\log(p)}}{\sqrt{n\rho}}\ ,\]
which is  independent of $m^*$, unlike the condition $\tau \gtrsim \sqrt{  \frac{K\vee \log p }{m^*n}   }+     \frac{\log(p)\vee K }{n}$ for PECOK. This unpleasant feature is induced by the inflation of $\|\widetilde{\Sigma}-AC^*A^t\|_{op}$ with $\alpha$. 
Again, this weakness in the guarantees may be an artifact of the proof, which relies on the Davis-Kahan inequality for controlling the alignment between the sample eigenvectors associated with the $K$ largest eigenvalues and their population counterpart. 
\end{itemize}

\noindent All the results of this section are proved in Section \ref{sec:proof:CSC} of the supplement \cite{SUPPL18}.

\section{Approximate $G$-block covariance  models} \label{sec:approx}

In the previous sections, we have proved that under some separation conditions, COD and PECOK procedures are able to exactly recover the partition $G^*$. However, in practical situations, the separation conditions may not be met. Besides, if the entries of $\Sigma$ have been modified by an infinitesimal perturbation, then the corresponding partition $G^*$ would consist of $p$ singletons. 

As a consequence, it may be more realistic and more appealing from a practical point of view to look for a partition $G[K]$ with $K< |G^*|$ groups such that $\Sigma$ is close to a matrix of the form $ACA^t+ \Gamma$ where $\Gamma$ is diagonal and $A$ is associated to $G[K]$. This is equivalent to considering a decomposition $\Sigma=ACA^t+ \Gamma$ with $\Gamma$ non-diagonal, where the non-diagonal entries of $\Gamma$ are small. In the sequence, we write $R = \Gamma - \text{Diag}(\Gamma)$ for the matrix of the off-diagonal elements of $\Gamma$ and $D=\text{Diag}(\Gamma)$ for the diagonal matrix given by the diagonal of $\Gamma$.

In the next subsection, we discuss under which conditions the partition $G[K]$ is identifiable and then, we prove that  COD and PECOK are able to recover these partitions.

 \subsection{Identifiability of approximate $G$-block covariance models}\label{identifa}

When $\Gamma$ is allowed to be  not exactly equal to a diagonal matrix, we encounter a further identifiability issue, as a generic matrix $\Sigma$ may admit many decompositions  $\Sigma= ACA^t+\Gamma$. In fact, such a decomposition holds for any membership matrix $A$ and any matrix $C$ if we define $\Gamma= \Sigma - A C A^t$. 
So we need to specify the kind of decomposition that we are looking for.
For $K$ being fixed, we would like to consider the partition $G$ with $K$ clusters that maximizes  the distance between goups (e.g. $\MCOD(\Sigma,G)$)  while having the smallest possible  noise  term $|R|_{\infty}$. Unfortunately, such a partition $G$ does not necessarily exist and is not necessarily unique. Let us illustrate this situation with a simple example.

\noindent 
{\bf Example.} Assume that $\Sigma$ is given by
$\Sigma= \begin{bmatrix} 2r & 0 &0 \\ 0 & 2r &0 \\ 0& 0 & 2r\end{bmatrix} + I_p$,
with $r>0$, with the convention that  each entry corresponds to a block of size $2$. Considering partitions with $2$ groups and allowing $\Gamma$ to be non diagonal, we can decompose $\Sigma$ using different partitions.  For instance 
$$\Sigma = \underbrace{\begin{bmatrix} 2r & 0 & 0\\ 0&  r & r\\ 0 & r & r\end{bmatrix}}_{=A_{1}C_{1}A_{1}^t}+\underbrace{\begin{bmatrix} 0 & 0 & 0\\ 0 & r & -r\\0 & -r & r\end{bmatrix}+I_p}_{= \Gamma_1}=\underbrace{\begin{bmatrix} r & r  &0\\ r & r &0\\ 0 & 0  & 2r \end{bmatrix}}_{=A_{2}C_{2}A_{2}^{t}}+\underbrace{\begin{bmatrix} r & -r& 0 \\ -r & r&0 \\ 0 & 0 & 0\end{bmatrix}+I_p}_{= \Gamma_2}.$$
 Importantly, the two decompositions correspond to two different partitions $G_1$ and $G_2$ and both decompositions  have $|R_i|_{\infty}=r$ and 
  $\MCOD(\Sigma,G_i) = 2r=2|R|_{\infty}$, for $i=1,2$. In addition, no decomposition $\Sigma=ACA^t +D +R$ with associated  partition in $2$ groups, satisfies $\MCOD(\Sigma,G)>2r$ or $|R|_{\infty}<r$. As a consequence, there is no satisfying way to define a unique partition maximizing $\MCOD(\Sigma,G)$, while having $|R|_{\infty}$ as small as possible. 
 We show below that the cutoff $\MCOD(\Sigma,G)>2|R|_{\infty}$ is actually sufficient for partition identifiability. 

  For this, let us define 
$\mathcal{P}_j(\Sigma,K)$, $j = \{1, 2\}$ as the set of quadruplets ${(A,C,D,R)}$ such that $ \Sigma=ACA^t+ D+R$, with $A$ a membership matrix associated to a partition $G$ with $K$ groups with $\min_{k}|G_{K}|\geq j$, and  $D$ and $R$ defined as above. Hence $\mathcal{P}_1$ corresponds to partitions without restrictions on the minimum group size. For instance, singletons are allowed. In contrast   $\mathcal{P}_2$ only contains partitions without singletons. We define
\begin{align*}
 \rho_1(\Sigma,K)&=\max\ac{\MCOD(\Sigma,G) / |R|_{\infty}: (A,C,D,R)\in \mathcal{P}_1(\Sigma,K)\ \textrm{and}\ G\ \textrm{associated to}\ A},\\
 \rho_2(\Sigma,K)&=\max\ac{\Delta(C) / |R|_{\infty}: (A,C,D,R)\in \mathcal{P}_2(\Sigma,K)}.
\end{align*}
We view $\rho_1$ and $\rho_2$ as respective  measures of ``purity" of the block structure of $\Sigma$.

\begin{prp}\label{prp:identifR}~\
\begin{enumerate}
\item[(i)] Assume that $\rho_1(\Sigma,K)>2$. Then, there exists a unique partition $G$  such that there exists a decomposition  $\Sigma=ACA^t+ \Gamma$, with $A$ associated to $G$ and $\MCOD (\Sigma,G)> 2 |R|_{\infty}$. We denote by $G_{1}[K]$ this partition.
\item[(ii)] Assume that $\rho_2(\Sigma,K)>8$. Then, there exists a
 unique partition $G$ with $\min_{k}|G_{k}|\geq 2$, such that there exists a decomposition  $\Sigma=ACA^t+ \Gamma$, with $A$ associated to $G$ and $\Delta(C) > 8 |R|_{\infty}$.  We denote by $G_{2}[K]$ this partition.
 \item[(iii)] In addition, if both $\rho_1(\Sigma,K)>2$ and $\rho_2(\Sigma,K)>8$, then $G_{1}[K]=G_{2}[K]$.
\end{enumerate}
\end{prp}
The conditions $\rho_1(\Sigma,K)>2$ and  $\rho_2(\Sigma,K)>8$ are minimal for defining uniquely the partition $G_1[K]$. For $\rho_1$, this has been illustrated in the example above the proposition. For $\rho_2$, we provide a counter example when $\rho_2(\Sigma,K)=8$ in Section \ref{sec:examples:Sigma} of the supplement \cite{SUPPL18}.  The proof of Proposition \ref{prp:identifR}. 
is given in Section \ref{sec:proof:prp:identifR} of \cite{SUPPL18}.

 The conclusion of Proposition \ref{prp:identifR}  does essentially revert to that of Proposition \ref{identif} of Section \ref{sec:model} as soon as $|R|_{\infty}$ is small enough respective to the cluster separation sizes.
Denoting $K^*$ the number of groups of $G^*$, we observe that $G_1[K^*]= G^*$ and $G_2[K^*]= G^*$ if $m^*\geq 2$. Besides, $\rho_1(\Sigma,K)= \rho_2(\Sigma,K)=0$ for $K> K^*$. For $K< K^*$ and when $G_1[K]$ (resp. $G_2[K]$) are well defined, then the partition $G_1[K]$ (resp. $G_2[K]$) is coarser than $G^*$. In other words, $G_1[K]$ is derived from $G^*$ by merging groups $G^*_k$ thereby increasing $MCOD(\Sigma,G)$ (resp. $\Delta(C)$) while requiring $|R|_{\infty}$ to be small enough.

 We point out that, in general, there is no unique decomposition $\Sigma=ACA^t+\Gamma$ with $A$ associated to $G_{2}[K]$, even when $\min_{k}|G_{2}[K]_{k}|\geq 2$. Actually, it can be possible to change some entries of $C$ and $R$, while keeping $C+R$, $\Delta(C)$ and $|R|_{\infty}$ unchanged.

   \subsection{The  COD algorithm for approximate $G$-block covariance models}   \label{SEC:COD:APPROX}
      We show below that the COD algorithm is still applicable if $\Sigma$  has small departures from a block structure. 
We set $ \lambda_{min}(\Sigma)$ for the smallest eigenvalue of $\Sigma$.

  \begin{thm}\label{prop:approxCOD}
Under the distributional Assumption 1, there exist numerical constants $c_{1},c_{2}>0$ such that the following holds for all $\alpha \geq c_1L^2 \sqrt{ \frac {\log p }{n}}$. If, for some partition $G$ and  decomposition $\Sigma= A CA^t+ R+ D$, we have
\beq\label{eq:approximation_cod_g}
|R|_{\infty} \leq  \frac{\lambda_{min}(\Sigma)}{2\sqrt{2}}\alpha \quad  \text{ and }
\quad \text{MCOD} (\Sigma,G) > 3\alpha  |\Sigma|_{\infty}\ , 
\eeq
then COD  recovers $G$ with probability higher than  $1-c_{2}/p$.
\end{thm}
The proof is given in Section \ref{sec:proof:cod} of the supplement \cite{SUPPL18}. If $G$ satisfies the assumptions of Theorem \ref{prop:approxCOD}, then it follows from  Proposition~\ref{prp:identifR} that $G=G_1[K]$ for some $K>0$. First, consider the situation where   the tuning parameter $\alpha$ is chosen to be of the order $\sqrt{\log(p)/n}$. If $\MCOD(\Sigma,G^*)\geq 3\alpha |\Sigma|_{\infty}$, then  COD selects $G^*$ with high probability. If $\MCOD(\Sigma,G^*)$ is smaller than this threshold, then no procedure is able to recover $G^*$ with high probability (Theorem \ref{thm:tlbM}). Nevertheless, COD is able to recover  a coarser  partition  $G_1[K]$ whose corresponding  MCOD metric $\MCOD(\Sigma,G)$ is higher than the threshold $3\alpha |\Sigma|_{\infty}$  and whose matrix $R$ is small enough. For larger $\alpha$, then COD recovers a coarser partition $G$ (corresponding to $G_1[K]$ with a smaller $K$) whose corresponding approximation $|R|_{\infty}$ is allowed to be larger. 

  \subsection{The PECOK algorithm for approximate $G$-block covariance models}\label{Pecok:approximate}

In this subsection, we investigate the behavior of PECOK under the approximate $G$-block models. The number $K$ of groups being fixed, we assume that $\rho_2(\Sigma,K)>8$ so that $G_2[K]$  is well defined. We shall prove that PECOK recovers $G_2[K]$ with high probability. By abusing the notation, we denote in this subsection $G^*$ for the target partition $G_2[K]$, $B^*$ for the associated partnership matrix and $(A,C^*,D,R)\in \cP_2(\Sigma,K)$ any decomposition of $\Sigma$ maximizing $\Delta(C)/|R|_{\infty}$.

Similarly to  Proposition \ref{prop:sdp-pop}, we first provide sufficient conditions on $C^*$ under which  a population version of PECOK can recover the true partition. 
\begin{prp}\label{prop:sdp-pop-approx}
If, $\Delta(C^*)> {7|D|_{V}+2 \|R\|_{op}\over m}+3|R|_{\infty}$\ , then  $B^*=\argmin_{B\in\mathcal{C}} \langle \Sigma, B\rangle$. 
\end{prp}

\begin{cor}\label{cor:sdp-pop-approx}
If  $\Delta(C^*)> 3|R|_{\infty}+{2 \|R\|_{op}\over m}$, 
then
$B^*=\argmin_{B\in\mathcal{C}} \langle \Sigma-D, B\rangle$. 
\end{cor}
In contrast to the exact $G$-block model, the cluster distance $\Delta(C^*)$ now needs to be larger than $|R|_{\infty}$ for the population version to recover the true partition.  The $|R|_{\infty}$ condition is fact necessary as discussed in subsection \ref{identifa}. 
In comparison to the necessary conditions discussed in subsection \ref{identifa},  there is an additional $\|R\|_{op}/m$ term. The proofs are given in Section \ref{A2} in the supplementary material \cite{SUPPL18}. \\

We now examine the behavior of PECOK when we specify  the estimator  $\widehat{\Gamma}$ to be as in \eqref{eq:estim:gamma2}.  Note that in this approximate block covariance setting, the diagonal estimator $\widehat{\Gamma}$ is in fact an estimator of the diagonal matrix $D$.  In order to derive deviation bounds for our estimator $\widehat \Gamma$, we need the following diagonal dominance assumption.
\\

 \noindent 
{\bf Assumption 2}: (diagonal dominance of  $\Gamma$) The matrix $\Gamma=D+R$ fulfills
\begin{equation}\label{assume:dominance}
\Gamma_{aa}\geq 3\max_{c:c\neq a} |\Gamma_{ac}|\quad \textrm{(or equivalently}\quad D_{aa}\geq 3\max_{c:c\neq a} |R_{ac}|\textrm{).}
\end{equation}

The next theorem states that PECOK estimator $\widehat{B}$ recovers the groups under similar conditions to that of Theorem \ref{THM:CONSISTENCY} if $R$ is small enough. 
The proof is given in Section \ref{T4.7} of the supplement \cite{SUPPL18}.

\begin{thm}\label{prop:consistency-general-gamma-hat}
There exist $c_1$, $c_2$, $c_L$, $c'_{L}$ four positive constants such that the following holds.  Under Assumptions 1 and 2, and when 
$L^4\log(p) \leq c_1 n$ and 
\begin{equation}\label{eq:condition_R} 
|R|_{\infty}+ \frac{\sqrt{|R|_{\infty}|D|_{\infty}}+\|R\|_{op}}{m} \leq c_L\|\Gamma\|_{op}\left\{\sqrt{  \frac{\log p}{mn}   }+    \sqrt{\frac{p}{nm^2}} + \frac{\log(p)}{n}+ \frac{p}{nm}\right\} 
\end{equation}
we have $\widehat{B} = B^*$  and $\widehat{G}=G^*$, with probability higher than $1 - c_2/p$, as soon as 
\begin{equation}\label{eq:condition_assumption2_gamma_hat}
 \Delta(C^*) \geq c'_L \left[\|\Gamma\|_{op}\left\{\sqrt{  \frac{\log p}{mn}   }+    \sqrt{\frac{p}{nm^2}} + \frac{\log(p)}{n}+ \frac{p}{nm}\right\}   \right]\ ,
\end{equation}
\end{thm}
So, as a long as $|R|_{\infty}$ and $\|R\|_{op}$ are small enough so that \eqref{eq:condition_R} are satisfied, 
  PECOK algorithm will correctly  identify the target partition $G^*$ at the $\Delta$-(near) optimal minimax level (\ref{eq:condition_assumption2_gamma_hat}). 
A counterpart of Theorem \ref{prop:consistency-general-gamma-hat} for Assumption 1-bis is provided in  Section \ref{T4.7} of the supplement \cite{SUPPL18}.


\section{Data analysis}\label{sec:data}
Using functional MRI data, \cite{power2011functional} found that
the human brain putative  areas  are organized into clusters, sometimes
referred to as networks or functional systems. We use a publicly available fMRI dataset
to illustrate the clusters recovered by different methods. The dataset was originally
published in \cite{XueG08} and is publicly available from Open fMRI (\url{https://openfmri.org/data-sets})
under the accession number ds000007. We will focus on analyzing two
scan sessions from subject 1 under a visual-motor stop/go task (task
1). Before performing the analysis, we follow the preprocessing
steps suggested by \cite{XueG08}, and we   follow \cite{power2011functional} to subsample the whole brain data using  $p=264$ putative  areas, see Section~\ref{app:fmri} of the supplementary materials \cite{SUPPL18}
   for details. This subject was also scanned in two separate sessions, and each session yielded  $n=180$ samples for each putative area. 

\begin{figure}
\protect\caption{\label{figfmriloss} (a) Plot of the coordinates of the largest COD cluster overplayed over a standard brain template.  The coordinates are shown as red balls.  (b) Comparison of \textsc{COD},  \textsc{COD-cc}, \textsc{PECOK},  K-means,  HC, and SC   using the Frobenius prediction loss criterion (\ref{eq:fmriloss})
where the groups are estimated by these methods respectively.   
}
 
\centering{}\subfloat[A brain cluster by \textsc{COD}]{\protect\includegraphics[scale=0.52]{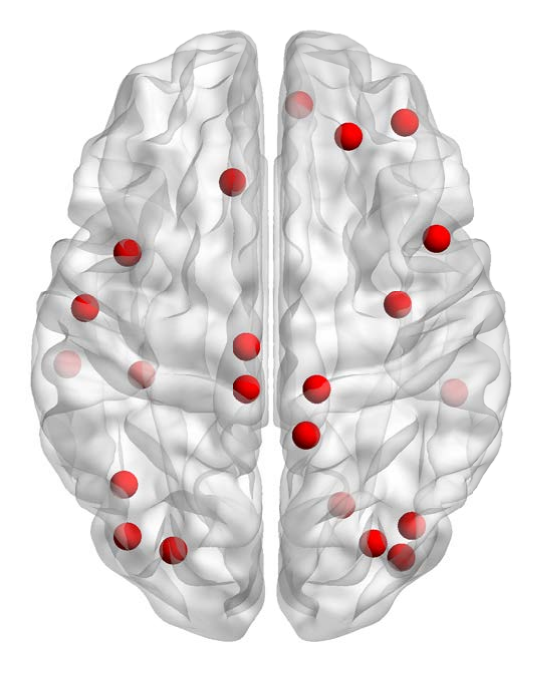}}\hfill{}\subfloat[Prediction loss]{\protect\includegraphics[scale=0.4]{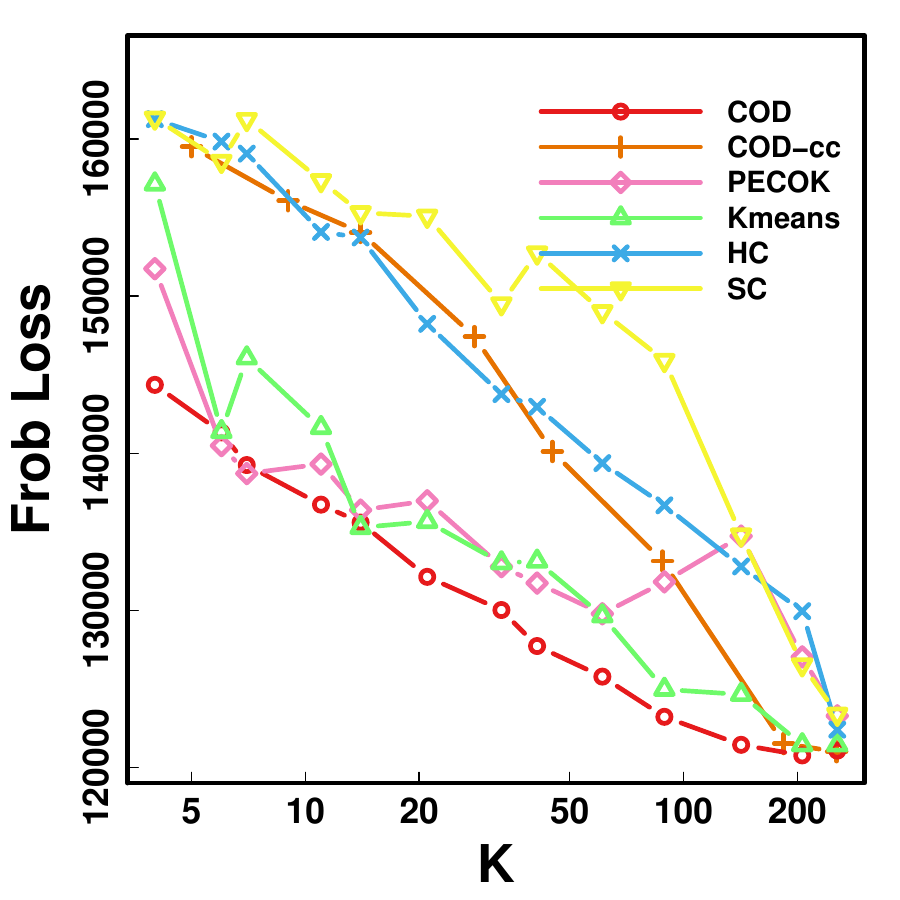}}

\end{figure}

We apply our data-splitting approach described in Section~\ref{sec:cvcod} to these two session data. Using the first scan session data only, we first estimate  $\hat{G}$ using \textsc{COD}  and \textsc{COD-cc}  on  a fine grid of $\alpha = c \sqrt{\log(p) /n }$ where $c=0.5, 0.6, \dotsc, 3$. For a fair comparison, we set $K$ in \textsc{PECOK} to be the same as the resulting   $K$'s found by \textsc{COD}. We then use the second session data to evaluate the loss $\mathcal{H}(G)$ given in Section~\ref{sec:cvcod}.    Among our methods (\textsc{COD}, \textsc{COD-cc}, and \textsc{PECOK}), \textsc{COD} yields the smallest loss when $K=142$. We thus first focus on illustrating the COD clusters here.   Table~\ref{tab:mni} lists the largest cluster of  putative areas  recovered by COD and their functional classification based on prior knowledge. Most of these  areas  are classified to be related to visual, motor, and task functioning, which is consistent with the implication of our experimental task that requires the subject to perform  motor responses based on visual stimuli.  Figure~\ref{figfmriloss}(a) plots the locations of these coordinates on a standard  brain template.  It shows that our COD cluster appears to come mostly from approximately symmetric locations from the left and right hemisphere, though we do not enforce this brain function symmetry in our algorithm.  Note that  the original coordinates in \cite{power2011functional} are  not sampled with exact symmetry from both hemispheres of the brain, and thus we do not expect exact symmetric locations in the resulting clusters based on these coordinates. 

Because there are no gold standards for  partitioning the brain, we follow common practice and  use a prediction
criterion  to further compare  the clustering performance of  different methods.  For a fair comparison, we also estimate $\hat{G}$ using K-means, HC,  and  spectral clustering on the same resulting   $K$'s found by \textsc{COD}.    The prediction criterion is as follows. 
We first compute the covariance
matrices $\hat{S}_{1}$ and $\hat{S}_{2}$ from the first and second
session data respectively. For a grouping estimate $\hat{G}$, we
use the following loss to evaluate its performance 
\begin{equation}
\left\Vert \hat{S}_{2}-\varUpsilon\left(\hat{S}_{1},\hat{G}\right)\right\Vert _{F}.\label{eq:fmriloss}
\end{equation}
where block averaging operator   $\varUpsilon\left(R,G\right)$  produces a $G$-block structured matrix based on $\hat{G}$.   For any
$a\in G_{k}$ and $b\in G_{k^{\prime}}$, the output matrix entry
$\left[\varUpsilon\left(R,G\right)\right]_{ab}$ is given by
\[
\left[\varUpsilon\left(R,G\right)\right]_{ab}=\begin{cases}
\left|G_{k}\right|^{-1}\left(\left|G_{k}\right|-1\right)^{-1}\sum_{i,j\in G_{k},i\ne j}R_{ij} & \mbox{if \ensuremath{a\ne b} and \ensuremath{k=k^{\prime}}}\\
\left|G_{k}\right|^{-1}\left|G_{k^{\prime}}\right|^{-1}\sum_{i\in G_{k},j\in G_{k^{\prime}}}R_{ij} & \mbox{if \ensuremath{a\ne b} and \ensuremath{k\ne k^{\prime}}}\\
1 & \mbox{if }a=b.
\end{cases}
\]
In essence, this operator smooths  over the matrix entries   with indices in 
the same group, and one may expect that such smoothing over variables in the true cluster   will reduce the loss \eqref{eq:fmriloss} while smoothing over  different clusters will increase the loss.  

Figure \ref{figfmriloss}(b) compares the  prediction loss values 
under different group sizes for each method. This shows that our  data-splitting approach for  \textsc{COD} indeed selects a value $K = 142$ that is immediately next to a slightly larger one ($K=206$), the latter having  the smallest prediction loss, near the bottom plateau. However, the differences are almost negligible. This  suggests that our data-splitting criterion, which comes with theoretical guarantees, also provides good prediction performance in this real data example, while  selecting  a slightly smaller $K$, as desired, since this makes the resulting clusters easier to describe and interpret. 

Regardless of the choice of $K$ or $\alpha$, Figure \ref{figfmriloss}(b) also shows that COD almost always yields the smallest prediction loss for a wide range of $K$, while PECOK does slightly better when $K$ is between 5 and 10. Though \textsc{COD-cc} has large losses for medium or small $K$, its performance is very close to the best performer  \textsc{COD} near $K=146$.  Kmeans in this example is the closest competing method, while the other two methods (HC and SC) yield larger losses across the choices of $K$. 

\begin{table}
\protect\caption{MNI coordinates (x, y, z, in mm) of the largest \textsc{COD} group and their
functioning classification.}\label{tab:mni}

\begin{centering}
\begin{tabular}{ccccccccc}
\hline 
X & Y & Z & Function &  & X & Y & Z & Function\tabularnewline
\hline 
\hline 
40 & -72 & 14 & visual & &  -7 & -21 & 65 & motor\tabularnewline
\hline 
-28 & -79 & 19 & visual & & -7 & -33 & 72 & motor\tabularnewline
\hline 
20 & -66 & 2 & visual & & 13 & -33 & 75 & motor\tabularnewline
\hline 
29 & -77 & 25 & visual & & 10 & -46 & 73 & motor\tabularnewline
\hline 
37 & -81 & 1 & visual & & 36 & -9 & 14 & motor\tabularnewline
\hline 
47 & 10 & 33 & task & & -53 & -10 & 24 & motor\tabularnewline
\hline 
-41 & 6 & 33 & task & & -37 & -29 & -26 & uncertain\tabularnewline
\hline 
38 & 43 & 15 & task & & 52 & -34 & -27 & uncertain\tabularnewline
\hline 
-41 & -75 & 26 & default &  & -58 & -26 & -15 & uncertain\tabularnewline
\hline 
8 & 48 & -15 & default & & -42 & -60 & -9 & attention\tabularnewline
\hline 
22 & 39 & 39 & default & & -11 & 26 & 25 & saliency\tabularnewline
\hline 
\end{tabular}
\par\end{centering}
\end{table}

%

\section{Discussion}\label{sec:discussion}
In this section, we discuss some related models and give an overall recommendation on the usage of our methods.

\subsection{Comparison with Stochastic Block Model} 

The problem of variable clustering that we consider in this work is fundamentally different from that of variable clustering from {\it network data}. The latter, especially in the context of the  Stochastic Block Model (SBM), has received a large amount of attention over the past years, for instance  \cite{guedon2014community,LeiRinaldo,chen2014statistical,lei2014generic,abbe2015community,mossel2014consistency,le2014optimization}.  The most important difference stems from the nature of the data:  the data analyzed via the SBM is a $p\times p$ binary  matrix $\bA$, called the adjacency matrix,  with entries assumed to have been generated as independent Bernoulli random variables; its expected value is assumed to have a block structure. In contrast,  the data matrix $\bX$ generated from a $G$-block covariance is a $n\times p$ matrix with real entries, and rows viewed as i.i.d copies of a $p$-dimensional vector $X$ with mean zero and dependent entries. The covariance matrix $\Sigma$ of $X$ is assumed to have (up to the diagonal) a  block structure. 

\medskip 

{\bf Need for a correction}. Even though the analysis of the methods  in our setting would differ from the SBM setting, we could have applied available clustering procedures tailored for SBMs  to the empirical covariance matrix $\widehat{\Sigma}= \bX^t\bX/n$ by treating  it as some sort of weighted adjacency matrix. 
 It turns out that applying verbatim the spectral clustering procedure of Lei and Rinaldo~\cite{LeiRinaldo} or the SDP such as the ones  in \cite{amini} would lead to poor results. The main reason for this is that, in our setting, we need to {\bf correct} both the spectral algorithm and the SDP to recover the correct  clusters (Section \ref{sec:pecok}). Second, the
 SDPs studied in the SBM context (such as those of \cite{amini}) do not handle properly groups with different and unknown sizes, contrary to our SDP. To the best of our knowledge, our SDP (without correction) has only been independently studied by Mixon et al.~\cite{ward} in the context of Gaussian mixtures.

 \medskip

{\bf Analysis of the SDP}. As for the mathematical arguments, our analysis of the SDP in our on covariance-type  model differs from that in mean-type models partly because of the  the presence of non-trivial cross-product terms. Instead of relying on dual
certificates arguments as in other work such as~\cite{2015arXiv150705605P}, we directly investigate the primal problem and  combine different duality-norm bounds. The crucial step is the Lemma \ref{lem:clef} in the supplementary material \cite{SUPPL18} which allows to control the Frobenius inner product by a (unusual) combination of $\ell^{1}$ and spectral control. In our opinion,  our approach is more transparent than dual certificates techniques, especially in the presence of a correction $\widehat \Gamma$ and allows for the attainment of  optimal convergence rates.

\subsection{Extension to other Models}
The general strategy of correcting a convex relaxation of $K$-means can be applied to other models. In \cite{MartinNIPS}, one of the authors has adapted  the PECOK algorithm  to the clustering problem of  mixture of subGaussian distributions. In particular, in the high-dimensional setting where the correction plays a key role, \cite{MartinNIPS} obtains  sharper separation conditions dependencies than in state-of-the-art clustering procedures~\cite{ward}.  Extensions to model-based overlapping clustering are beyond the scope of this paper, but we refer to \cite{BBNW-overlap2017} for recent results.

\subsection{Practical recommendations}\label{practical} 
 Based on our extensive simulation studies, we conclude this section with general recommendations on the usage of our proposed algorithms. 

If $p$ is moderate in size,  and  if there are reasons to believe that no singletons exist in a particular application, or if they have been removed in a pre-processing step, we recommend the usage of the PECOK algorithm, which is  numerically superior to existing methods: exact recovery can be reached for relatively small  sample sizes.  COD is also very competitive, but requires a slightly larger sample size to reach the same performance as PECOK.  The constraint on the size of $p$ reflects the existing computational limits  in  state-of-the art algorithms for SDP, not the statistical capabilities of the procedure, the theoretical analysis of which being one of the foci of this work. 

If $p$ is large, we recommend COD-type algorithms. Since COD is optimization-free, it scales very well with $p$, and only requires a moderate sample size to reach exact cluster recovery.  Moreover, COD adapts very well to data that contains singletons and, more generally, to data that is expected to have many inhomogeneous clusters.

\appendix

\section*{Acknowledgements}
We thank the editors and anonymous reviewers for their helpful suggestions. 
We thank Andrea Montanari for pointing to us the reference \cite{ward}.  
The project is partially supported by the CNRS PICS grant HighClust. Christophe Giraud is partially supported by the LabEx LMH, ANR-11-LABX-0056-LMH. Martin Royer is supported by an IDEX Paris-Saclay IDI grant, ANR-11-IDEX-0003-02. Xi Luo is partially supported  by NSF-DMS 1557467, NIH R01EB022911, P01AA019072, P20GM103645, P30AI042853, and S10OD016366. Florentina Bunea is partially supported by NSF-DMS 1712709.

\begin{supplement}[id=suppA]
  \sname{Supplement to}
  \stitle{``Model assisted variable clustering: minimax-optimal recovery and algorithms"}
  \slink[doi]{10.1214/00-AOASXXXXSUPP}
  \sdatatype{.pdf}" 
  \sdescription{This supplement contains:  proofs of  the theoretical results,  the  simulation results, and additional supporting information regarding the data analysis.}
\end{supplement}

\bibliography{biblio}
\bibliographystyle{plain}

\includepdf[pages=1-last]{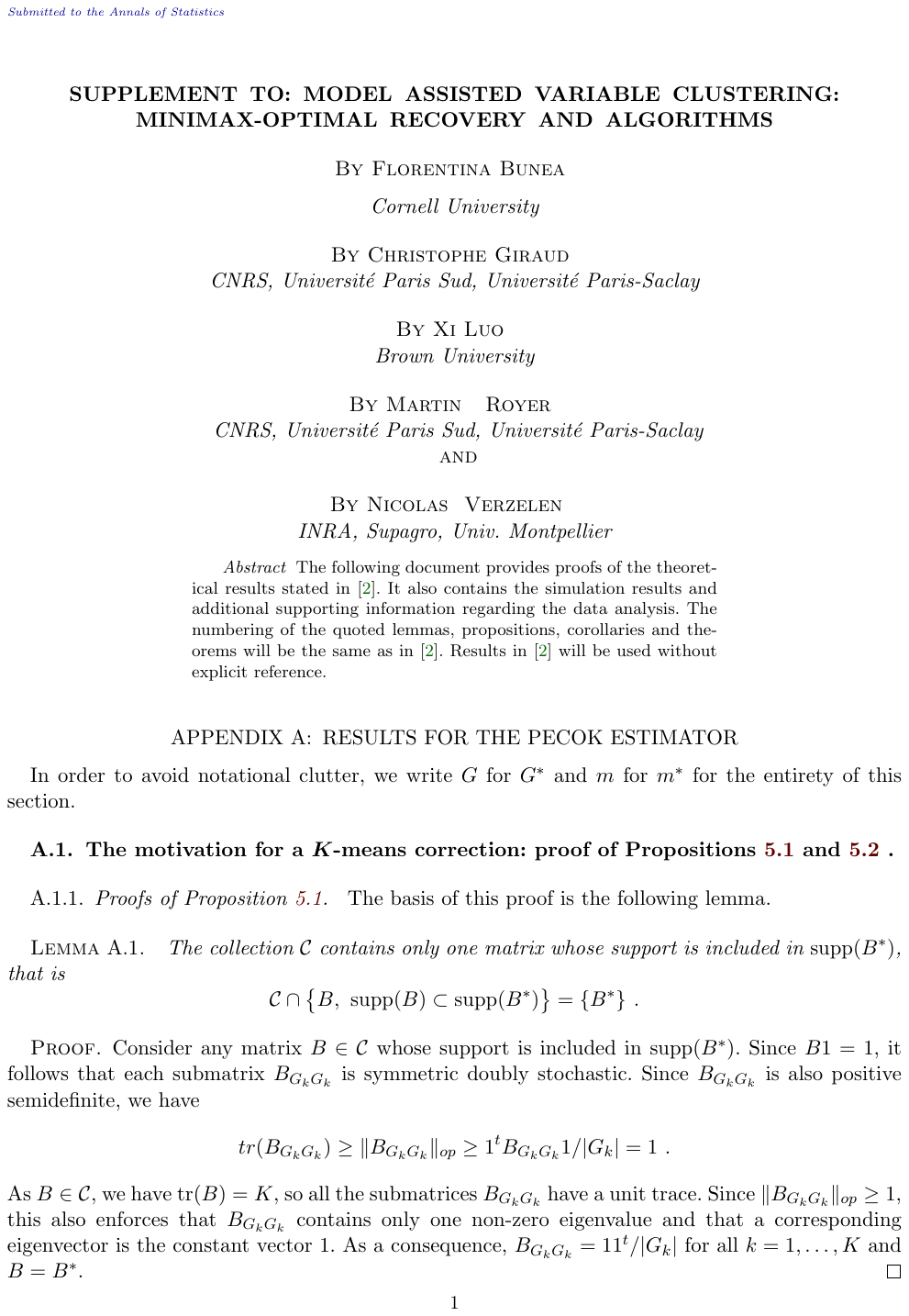}

\end{document}